\documentclass[12pt,a4paper]{article}
\pdfoutput=1
\usepackage{jheppub}

\usepackage{bm}
\usepackage{amsmath}
\usepackage{amssymb}
\usepackage{ifsym}
\usepackage{amsthm}
\usepackage{amsfonts}
\usepackage{ccaption}
\usepackage{booktabs}
\usepackage{mathrsfs}

\makeatletter
\@addtoreset{equation}{section}

\makeatletter
\renewcommand\section{\@startsection {section}{1}{\z@}%
                                   {-3.5ex \@plus -1ex \@minus -.2ex}
                                   {2.3ex \@plus.2ex}%
                                   {\normalfont\large\bfseries}}
\renewcommand\subsection{\@startsection{subsection}{2}{\z@}%
                                     {-3.25ex\@plus -1ex \@minus -.2ex}%
                                     {1.5ex \@plus .2ex}%
                                     {\normalfont\bfseries}}

  \captionnamefont{\bfseries}
  \captiontitlefont{\small\sffamily}
  \captiondelim{: }
  \hangcaption

\def\sect#1{\S\ref{#1}}
\def\fig#1{Fig.\,\ref{#1}}
\def\req#1{(\ref{#1})}

\definecolor{point}{rgb}{0.7,0.,0.3}

\def\AC{AdS/CFT}
\def\GG{gauge/gravity}

\def\GR{general relativity}
\def\GT{gauge theory}
\def\SYM{super-Yang-Mills}
\def\STY{string theory}
\def\QG{quantum gravity}
\def\Poinc{Poincar\' e}
\def\schw{Schwarzschild}
\def\RN{Reissner-Nordstr\o m}
\def\EE{entanglement entropy}

\def\RR{\mathbb{R}}
\def\gst{g_{\scriptscriptstyle {\rm s}}}
\def\gYM{g_{\scriptscriptstyle {\rm YM}}}
\def\GN{G_{\scriptscriptstyle {\rm N}}}
\def\lpl{\ell_{\scriptscriptstyle {\rm p}}}
\def\lst{\ell_{\scriptscriptstyle {\rm s}}}
\def\scri{\mathscr I}
\def\adss#1#2{AdS$_{#1} \times S^{#2}$}
\def\Rads{\ell}
\def\ph{\varphi}
\def\dda#1{\left( \frac{\partial}{\partial #1} \right)^{\! a}}
\def\rh{r_+}
\def\Sbh{S_{\scriptscriptstyle {\rm BH}}}
\def\Sgt{S_{\scriptscriptstyle {\rm YM}}}
\def\vev#1{\langle\, #1 \, \rangle}

\def\andeq{\qquad {\rm and} \qquad}
\def\dual{\simeq}

\title{The AdS/CFT Correspondence}

\author{Veronika E. Hubeny}

\affiliation{ Centre for Particle Theory \& Department of Mathematical Sciences, Durham University\\
Science Laboratories, South Road, Durham DH1 3LE, UK.}\emailAdd{veronika.hubeny@durham.ac.uk}

\abstract{
We give a brief review of the \AC\ correspondence, which posits the equivalence between a certain gravitational theory and a lower-dimensional non-gravitational one.  This remarkable duality, formulated in 1997, has sparked a vigorous research program which has gained in breadth over the years, with applications to many aspects of theoretical (and even experimental) physics, not least to general relativity and quantum gravity.  To put the \AC\ correspondence in historical context, we start by reviewing the relevant aspects of string theory (of which no prior knowledge is assumed).  We then develop the statement of the correspondence, and explain how the two sides of the duality map into each other.  Finally, we discuss the implications and applications of the correspondence, and indicate some of the current trends in this subject.  The presentation attempts to convey the main concepts in a simple and self-contained manner, relegating supplementary remarks to footnotes.} 

\begin{document}

\maketitle

\flushbottom
\renewcommand{\thefootnote}{\arabic{footnote}}
\newpage


\section{Prologue: thanks to black holes...}
\label{s:intro}

The beautiful theory of general relativity enjoyed many remarkable achievements over the last century and provided a crucial edifice for theoretical physics.  The conceptual revolution regarding the nature of space and time has popularized the theory to the extent that it would be hard to find a scientific-minded person who has not heard of Einstein's theory of gravity.
Yet, despite its universally recognized elegance, until recently general relativity has been used by a relatively small subset of the scientific community -- much smaller, for instance, than those using quantum mechanics or quantum field theory -- mostly\footnote{
Of course, there were also (comparatively much smaller numbers of) interested physicists in other fields, amongst them notably string theorists -- which, as we'll soon see, was pivotal in the present context.
} restricted to relativists, cosmologists, and a fraction of astrophysicists.  This is not so surprising; after all, the effects of general relativity are pretty negligible at the scales relevant in context of condensed matter physics or nuclear physics.  It would seem even more audacious to imagine that gravitation could play any interesting role deeper in the quantum world, say for quantum information theory.

Yet over the last decade, general relativity has percolated all of these fields!  Nowadays one can find many condensed matter physicists, nuclear physicists, quantum information theorists, and others, who are actively interested in general relativity, not just for idle curiosity, but as a crucial tool in their research.  What has nucleated this dramatic transition?  Naturally,  the overall scientific trend points toward multidisciplinarity as the value of cross-fertilization between different areas is becoming more widely appreciated -- but that is the effect rather than the cause.  The real game changer responsible for this transition was the \AC\ correspondence (now more generally known as a \GG\ duality), conjectured by Juan Maldacena in 1997 \cite{Maldacena:1997re}.  Indeed, as we will see,  the manner in which \AC\ related all these subjects to gravity could scarcely have been more spectacular: it turns out that it's not the weak curvature of our spacetime which comes into play when studying everyday systems, but the most gravitational object there is: the black hole!  Incredible as this may sound, it is even more amazing, though in retrospect also demystifying, that the relevant dynamical spacetime is a higher-dimensional one compared to the non-gravitational system it describes.  Indeed, the most striking feature of the \AC\ duality is its `holographic' nature: it posits full equivalence between certain  theories formulated in different numbers of dimensions.

We postpone a more detailed phrasing of the correspondence till \sect{s:AC}, after we have put it into historical context and explained the necessary ingredients in \sect{s:preAC}.\footnote{
Since the focus throughout this chapter is on explaining the concepts (rather than providing a comprehensive review or a historical account), we will attempt to keep references to a minimum,  based mainly on historical importance and suitability for the present audience.  For a more comprehensive list of references see the relevant reviews cited below.
}  
The \AC\ correspondence has been arrived at using the framework of string theory, and indeed engendered what might be called  
a scientific revolution within the subject. 
This new type of holographic duality not only provided a more complete formulation of the theory,  but also profoundly altered our view of the nature of spacetime:  the gravitational degrees of freedom emerge  as effective classical fields from  highly quantum gauge theory degrees of freedom.  This harks back to earlier expectations motivated by black hole thermodynamics, that spacetime arises as a coarse-grained effective description of some underlying microscopic theory, but with a new twist: the relevant description is lower-dimensional.  As we will see, black holes not only motivated this idea, but played a key role in both the derivation and subsequent applications of \AC.

Over the intervening years, the \AC\ correspondence has flowered into a vast and intricate subject; Maldecena's original paper \cite{Maldacena:1997re} alone now has well over 10,000 citations.  While the \AC\ duality as such has not been rigorously `proved' (partly because we do not yet have a complete independent definition of the \QG\  side of the correspondence), it has successfully withstood such an impressive array of highly nontrivial checks, that a vast majority of the community is now  fully convinced the duality holds.  Indeed, the evidence has mounted so rapidly, that the mindset soon changed from ``can it possibly be true?"  to ``what does it mean?" or ``how does it work?" and ``what else can we do with it?".
Hence regardless of its status as a theorem,  the utility of \AC\ has already been amply demonstrated by the consequent discoveries.  In fact, the resulting relations transcend the original construction; just as \GG\ duality can actually be formulated without any recourse to \STY, 
many of its implications in turn hold independently of  \AC.
Gratifyingly, the statement of equivalence between two sides of the correspondence stimulates information flow in both directions, teaching us important lessons about each side, as we review in \sect{s:appliedAC}.  
In particular, the \GG\ duality engendered  a number of fascinating observations about \GR\ itself.
Conversely,  \GR\ provides the best means hitherto 
to calculate certain interesting quantities within strongly coupled field theories -- which is the main reason why so many physicists are using \GR\ as a tool.  

Far-reaching as the implications of the \AC\ correspondence have turned out, the lessons of the correspondence are far from exhausted, with new surprises jumping at us from every corner -- indeed it is evident that we have barely seen the tip of the iceberg!  
Particularly enticing is the unquenched promise of `solving' \QG.  While this quest motivated the original explorations leading up to \AC\ as well as its subsequent developments, we still have a long journey ahead of us.  Already, the \GG\ duality has given us many crucial insights into \QG; in fact,  one review of the \GG\ duality \cite{Horowitz:2006ct} starts with the assertion that: {\it Hidden within every non-Abelian gauge theory, even within the weak and strong nuclear interactions, is a theory of quantum gravity.}  While part of the community is still immersed in mining `Applied \AC' to gain further insight into various systems of interest in condensed matter, nuclear, and particle physics, the overall focus is gradually shifting towards tackling the most mysterious aspects of the correspondence in view of unraveling \QG.  Quintessentially quantum notions such as entanglement are gaining in significance, and the links with quantum information theory are now being vigorously explored.  Where it will all lead we still don't know, but for now there are many exciting paths to follow.

\section{Strings and branes}
\label{s:preAC}

With the advent of the singularity theorems [reviewed in ch.8], it became manifest that classical \GR\ can break down quite generically: it ceases to provide a valid description of physics near curvature singularities.  The natural expectation of course is that quantum mechanical effects `resolve' the singularities, so that the dynamics remains well-defined throughout the entire evolution.  
However, since quantum mechanics and \GR\ by themselves are mutually incompatible, our description of the universe based solely on these would be not just incomplete but actually inconsistent.

In the ongoing quest for \QG, the most useful playground has been provided by black holes,\footnote{
We will consider spacetimes with well-defined future null infinity $\scri^+$ and use the standard definition of a black hole as the region inside the event horizon ${\cal H}^+ \equiv \partial I^-[\scri^+]$, i.e.\ one causally disconnected from the boundary.  The horizon is always a spacetime co-dimension 1 null surface, but in higher dimensions it may be extended in one or more directions, giving rise to black strings, black branes, etc.;  in the present context the term `black hole' refers to all such objects.
} for many reasons.  Obviously, typical black holes contain curvature singularities, which manifestly require new physics.  These might reveal underlying symmetries of the full theory 
(motivating for instance the `cosmological billiards' program started by \cite{belinskii1982general} based on universal Mixmaster-like behaviour near spacelike singularities), or suggest possibilities of how evolution could remain well-defined through a singularity (as happens  e.g.\ for certain timelike singularities in string theory),
but usually direct study suffers from insufficient calculational control.
Fortunately,
it turns out that we can gain important insight without being quite so ambitious: already the black hole horizon contains many profound clues,  as implied by black hole thermodynamics [reviewed in ch.10; see also \cite{Wald:1999vt}].  The relations between geometrical properties of the horizon (area and surface gravity) and thermodynamic quantities (entropy and temperature)  hint at a statistical mechanical origin. 
While Hawking \cite{hawking1975particle} famously explained the temperature in terms of semi-classical particle creation, understanding the origin of black hole entropy\footnote{
That entropy of a black hole should be proportional to its area has been  deduced by Bekenstein \cite{bekenstein1973black} based on various gedanken-experiments, but these did not give a clue to its underlying quantum nature.
}
 in terms of the corresponding number of  `microstates' has proved both more difficult and more rewarding.  Classically, the black hole `no-hair' theorems  imply that these microstates are not simply different  spacetime geometries that look like a black hole -- we need to go beyond \GR.
 Counting the black hole microstates has therefore provided an invaluable benchmark for putative theories of \QG.

To motivate whence one should seek guidance in this endeavor, recall that
in  units $c=1$, $\hbar =1$, the 4-dimensional Newton's constant is given by $\GN = \lpl^2$ where the Planck length $\lpl \approx 1.6 \times 10^{-33} \, cm$ specifies the length scale at which quantum effects become important. 
 Of course, direct access to such scales  is many orders of magnitude beyond the reach of accelerators, so the theory must be guided not by experiment but rather by internal consistency.  Fortuitously, this turns out to be an extremely stringent requirement.

Unfortunately, conventional field theory techniques are not well suited to quantizing general relativity directly:
 The fact that in 4 spacetime dimensions the dimensionless gravitational coupling  grows quadratically with energy gives rise to non-renormalizable perturbation theory: the divergences become uncontrollably worse at each order.  This indicates that some new physics has to kick in
to modify \GR\
  in the UV (meaning short distances or high energy scales).
An ingenious way to tame these divergences is to consider strings as the fundamental degrees of freedom (so that what we previously thought of as particles are simply different excitation modes of the string, a spatially extended 1-dimensional object), which effectively smears out the interactions.

This might at first sight seem rather fanciful, but it works!  More than that, it works spectacularly well.\footnote{
In fact, strings were originally introduced as a possible theory of the strong interactions already by the early 70s , since they nicely explained certain features of the hadron spectrum.  This program however dwindled with the advent of QCD,  until it was realized in the mid-70s that \STY\ is also a theory of gravity  \cite{scherk1974dual} (see e.g.\ \cite{Schwarz:2012zc} for a historical review and  \cite{Green:1987sp,Polchinski:1998rq} for the classic string theory textbooks).  It is amusing to note that two decades later the circle got  (almost) completed  via the \AC\ correspondence, wherein string theory turns out to be {\it dual} to a gauge theory akin to QCD.
}  
In a consistent Lorentz-invariant quantum theory, a closed string necessarily has a massless spin-two state, the graviton, whose long-wavelength interactions reproduce \GR.  Other states correspond to different particles: gauge bosons, fermions, etc; indeed \STY\ automatically incorporates the earlier ideas trying to explain the Standard Model such as grand unification, supersymmetry, and Kaluza-Klein theory.\footnote{
Consistency requires that the (weakly-coupled) theory is formulated in 10 dimensions, but these can be small (and therefore able to accommodate our spacetime looking 4-dimensional). The geometry of these `internal' dimensions then determines the 4-dimensional matter content.
}
In the course of trying to unravel \STY, it became evident that it in fact encompasses other `competing' constructions as well; indeed, within a few years it loomed into a vast edifice with incredibly rich structure which contains many essential features from other areas of both physics and mathematics.

Given that any string theory is a quantum theory which necessarily includes gravity, it naturally invites the examination of its consequences and implications for \QG.  Since spacetime plays a central role in \GR, the obvious question to ask is how does \STY\ give rise to the curved dynamical spacetime of \GR,  and what happens to the `stringy geometry' when the classical description breaks down.  
In the perturbative formulation valid at small string coupling $\gst$, 
 the spacetime coordinates specifying the position of the string appear as scalar fields in the 2-dimensional sigma model describing the dynamics of the string worldsheet, and
the spacetime metric then enters as a coupling constant.
But because the strength of gravitational interactions is governed by the string coupling -- the $d$-dimensional Newton's constant is $\GN = \gst^2 \, \lst^{d-2}$  in units of the string length $\lst$ --  it is difficult to directly access the most interesting, strongly gravitational, regime.  

Nevertheless, already at this level we encounter several intriguing surprises.  Since strings are extended objects, some spacetimes which are singular in \GR\ (for instance those with a timelike singularity akin to a conical one) appear regular in \STY.  Spacetime topology-changing transitions can likewise have a completely controlled, non-singular description.  Moreover, the so-called `T-duality' equates geometrically distinct spacetimes: because strings can have both momentum and winding modes around compact directions, a spacetime with a compact direction of size $R$ looks the same to strings as spacetime with the compact direction having size $\lst^2 / R$, which also implies that strings can't resolve distances shorter than the string scale $\lst$.  Indeed this idea is far more general  (known as mirror symmetry \cite{Greene:1991iv}), and exemplifies why spacetime geometry is not as fundamental as one might naively expect.

However, to learn about the more interesting regime where gravity is strong, we need to go beyond perturbation theory.  At first sight this looks extremely challenging, since generically a quantum theory at strong coupling has uncontrollably large fluctuations.  But remarkably  in string theory the situation is far tamer.  It turns out that string theories have a symmetry known as string duality: a strongly coupled limit of any given \STY\ is equivalent to a weakly coupled limit of another one.\footnote{
There are actually five perturbative 10-dimensional string theories, which differ from each other by the way the supersymmetry acts on the string and whether it contains just closed or both open and closed strings; in the following, when we say `string theory' we mean the `Type IIB string theory' (and similarly for its low-energy limit, supergravity).  There is also a sixth corner corresponding to an 11-dimensional theory called M-theory, wherein the size of the 11th dimension may be associated with the strength of the string coupling and whose fundamental excitations are 2-branes and 5-branes.}  
In other words, as we take the coupling large (i.e.\ in the $\gst \to \infty$ limit), the theory actually simplifies, and can be described perturbatively in some dual variables.
Though the intricate web of such relations was gradually built up through the early 90's, the crucial insight was provided by Polchinski  \cite{Polchinski:1995mt}, who showed that so-called Dirichlet-branes,\footnote{
Using arguments based on T-duality, the idea of D-branes was in fact introduced already in \cite{Dai:1989ua}, 
as extended objects which can couple consistently to strings (and thence conjectured to be in a sense made up of strings).  For a technical overview see e.g.\ \cite{Polchinski:1996na}.
}  
or D-branes for short, provide the  necessary charges\footnote{
Namely, the so-called Ramond-Ramond (RR) charges.  These are required by duality transformations, but cannot be carried by strings.  It was already known that they can be carried by black $p$-branes \cite{Horowitz:1991cd} (black holes  which are extended in $p$ spatial directions), which prompted the speculation that these black $p$-branes might be related to the dual variables. 
}
 required by the duality.
At weak coupling, D-branes behave as topological defects on which open strings can end, but in the full theory they are dynamical objects with mass scaling inversely with string coupling.  Hence at strong coupling they become light, providing the natural `fundamental' constituents of the theory.

The advent of string dualities immediately stimulated the program of black hole microstate counting.
   To explain this, let us take a short detour into specifying a few essential features of D-branes, which will pave the way towards motivating the \AC\ correspondence.
D-branes can be spatially
 extended in any number $p$ of dimensions;
   the world-volume of a  D$p$-brane is $(p+1)$-dimensional.  
  Massless modes of open strings ending on a D-brane describe the transverse fluctuations of the brane and give rise to gauge fields along the brane and their fermionic partners; for  $N$ coincident D-branes the low energy dynamics is then given by a $U(N)$ gauge theory.\footnote{
Every open string has two endpoints, each lying on one of the branes.
When the branes coincide the symmetry gets enhanced since the strings stretching between them become massless.
}  A stack of $N$ coincident D$p$-branes has $N$ units of $(p+1)$-form charge, and preserves half the supersymmetry.
 This has the happy consequence that the number of microstates of such a system  remains invariant under varying the string coupling $\gst$, and  the mass of such states does not receive quantum corrections.
To take advantage of this observation, we need to understand what happens to the system at strong coupling.

The amount by which  $N$ coincident D-branes backreact on the spacetime is governed by 
$\GN \, M \sim \gst^2 \, \frac{N}{\gst} =  \gst \, N$ in string units.  At sufficiently weak coupling, $\gst \, N \ll 1$, the branes live in nearly-flat 10-dimensional spacetime, whereas for $\gst \, N \gg 1$, the branes backreact strongly and source an extremal black brane geometry \cite{Horowitz:1991cd}, a 10-dimensional analog of the extremal \RN\ black hole, extended in $p$ spatial directions.\footnote{
More precisely, the geometry and causal structure depend on $p$; the horizon is regular for $p=3$ and for appropriate bound states of multiple branes, which requires the system to carry more than two types of charge.
For purposes of evaluating the entropy, given by quarter the horizon area {\it in Planck units}, the extra $p$ dimensions however do not play any role because the higher-dimensional Newton's constant is related to the lower-dimensional one by the same factor (namely the volume of the internal space) that relates the corresponding horizon areas. 
}
 In this regime, one can evaluate the horizon area
 and compare this with the statistical entropy obtained from counting the degeneracy of the system carrying the same charges at weak coupling.
This idea was realized in the seminal work of Strominger \& Vafa \cite{Strominger:1996sh}
which correctly counts the Bekenstein-Hawking entropy of a  5-dimensional 3-charge extremal black hole (a strongly-coupled description of a dimensionally reduced D1-D5 bound state carrying momentum).  Note that this is  a far greater achievement than merely matching one number: both the functional dependence on all the charges as well as the overall coefficient come out correctly without any additional input.  Since the weakly-coupled calculation looks nothing like the strongly coupled one, this nontrivial success suggests that such black holes are indeed `made of' D-branes.

This new understanding unleashed a flurry of activity which
reproduced the microscopic entropy of many other black holes described by more parameters, both in 4 and 5 dimensions and for other charges and angular momentum; see e.g.\ \cite{Horowitz:1996qd,Horowitz:1996rn} for early reviews and \cite{Dabholkar:2012zz} for a broader overview. 
One might expect that supersymmetry is essential, but in fact one can reproduce the entropy of some nonextremal black holes as well.  Although such black holes are quantum mechanically unstable due to Hawking evaporation, we can describe the Hawking process at weak coupling, as emission of closed strings from the brane.  More impressively still, the weakly-coupled D-brane description can even correctly reproduce the full radiation  spectrum of the black hole, including the grey-body factors due to the curved spacetime through which the Hawking quanta propagate  \cite{Maldacena:1996ix}.

As an aside, we should remark that sufficiently far from extremality we do lose control.  So the `simplest' black hole from the general relativistic standpoint and one which is perhaps closest to the hearts of most relativists, namely the \schw\ solution, still eludes the exact microstate counting.  Nevertheless, the scaling of entropy with area (as well as correct dependence on other charges) can be reproduced quite generally, by viewing the black hole at weak coupling as a bound state of strings and D-branes with the same conserved charges.  The relation between the two descriptions was established by Horowitz \& Polchinski    \cite{Horowitz:1996nw} and is known as the ``correspondence principle":
Consider an excited fundamental string.  As we increase the string coupling $\gst$, the string self-gravitates more strongly and shrinks, while simultaneously the Schwarzschild radius $\GN \, M$ increases.  At high enough coupling the string shrinks to within its own Schwarzschild radius, at which point the object is more appropriately described as a black hole.  Conversely, if we start with a black hole and decrease the coupling,  the horizon eventually becomes string size (which can however still be macroscopic in Planck units),  at which point we can no longer trust the classical black hole solution and the system is more appropriately described as an excited string.  Matching the masses\footnote{
In fact, this resolved a puzzle that previously impeded advancing the suggestion \cite{Susskind:1993ws} that black holes are really excited string states; namely, the respective entropies don't scale the same way with the mass:  $S_{string} \sim \lst \, M$ while (say in 4 dimensions) $S_{BH} \sim \GN \, M^2$.  The resolution is that both masses cannot be kept fixed simultaneously as we vary the string coupling.  Making them match at the transition when the black hole is string size ($\gst \sim (M \, \lst)^{-1/2}$) makes the entropies agree as well.  This agreement continues to hold more generally in all dimensions and in the presence of extra charges, angular momentum, etc.
} on the two sides of the transition, \cite{Horowitz:1996nw} found that the entropies also match, up to a constant of order unity which depends on the precise transition point. 

By 1996 string theory has made significant advances towards understanding the quantum aspects of black holes.  Cracking the black hole Information Paradox seemed to be just around the corner, and there were many tantalizing hints at a deeper structure underlying the many newly-discovered mysterious connections.  Though the former hope was rather too optimistic, little did we expect how substantial a progress the following year will see on the latter front.

\section{The \AC\ correspondence}
\label{s:AC}

In November 1997 Maldacena wrote his groundbreaking paper \cite{Maldacena:1997re}, conjecturing the relation which became known as the \AC\ correspondence.  In \sect{s:Malda} we will explain how Maldacena arrived at this remarkable conjecture, indicate its formulation and key features, and mention its immediate sociological impact.  Since the statement is perhaps as mystifying as it is profound, we pause in \sect{s:ACmodern} to give a more modern perspective which motivates the correspondence without any recourse \STY.  We then build up the basics of the \AC\ dictionary which simultaneously exemplify some of the initial checks of its validity in \sect{s:ACdictionary} and \sect{s:BHs}, the latter focusing on the important context of AdS black holes.  Finally, we  briefly mention several generalizations of \AC\ in \sect{s:generalizations}.
For further details, we refer the reader to the excellent early review by `MAGOO' \cite{Aharony:1999ti} (see also \cite{DHoker:2002aw,Polchinski:2010hw}).

\subsection{Maldacena's derivation}
\label{s:Malda}

The success in unraveling the D1-D5 system inspired analogous calculations for a D3-brane system, which provided the focal example of Maldacena's conjecture \cite{Maldacena:1997re}.
Consider $N$ coincident D3-branes (in type IIB string theory).
At weak coupling $\gst \, N \ll 1$, the branes live on a flat 10-dimensional spacetime and we have open strings 
ending on the D-branes
as well as closed strings propagating in the bulk.
On the other hand, at strong coupling $\gst \, N \gg 1$, the branes curve the spacetime substantially, sourcing the extremal black 3-brane geometry \cite{Horowitz:1991cd}:
\begin{equation}
ds^2 = f(r)^{-1/2} \,  \eta_{\mu\nu} \, dx^\mu \, dx^\nu 
+  f(r)^{1/2} \, \left( dr^2 + r^2 \, d\Omega_5^2 \right)
\ \ , \qquad
f(r) = 1+ \frac{4 \pi \, \gst \, N \, \lst^4}{r^4}
\label{e:ebb3}
\end{equation}	
where $x^\mu$ denote the 4 coordinates along the D3-brane worldvolume and $d\Omega_5^2$ is the metric of a unit $S^5$.  The solution is supported by a self-dual 5-form field strength, which has flux on the $S^5$.  

So far, we have specified two distinct regimes of $\gst N$ with no region of overlap.  Maldacena's insight was to consider decoupling the theory on the branes from gravity.  This can be achieved by taking a low-energy limit, which simplifies the physics enormously.    
On the one hand, the open string sector decouples from the rest of the theory, so that we end up with a 4-dimensional $SU(N)$ gauge theory (specifically super Yang-Mills) describing the dynamics of the branes.  On the other hand, in the black brane spacetime, this limit\footnote{
Actually, \cite{Maldacena:1997re} implemented this by starting with slightly-separated branes and taking the $\lst \to 0$ limit at fixed (large) $N$ and (small) $\gst$ while simultaneously bringing the branes together, keeping the open string mass fixed.  In practice, this can be accomplished by taking $r \to 0$ in \req{e:ebb3}. } focuses on the near-horizon geometry:  from the asymptotic viewpoint, any finite-energy excitation near the horizon will be strongly redshifted, while modes which propagate in the asymptotic region of \req{e:ebb3} decouple from the near-horizon region (since its cross section vanishes in this limit). 

Note that just as in the case of extremal \RN\ black hole,  the event horizon $r=0$ of the black brane solution  \req{e:ebb3} lies at infinite proper distance along spacelike geodesics; its embedding diagram has an infinite `throat'. The near-horizon geometry then has an enhanced symmetry, and simplifies to a direct product of a sphere and Anti-de-Sitter spacetime.  In case of 4-dimensional extremal \RN\ black hole, the near-horizon geometry is simply \adss22,  while in the present 10-dimensional case we have \adss55.  In particular, defining  $\Rads \equiv (4 \pi \, \gst \, N)^{1/4} \, \lst$ in \req{e:ebb3}, we see that  as $r \to 0$ (zooming near the horizon), 
$f(r)^{1/2} \to \Rads^2/r^2$, which obtains
\begin{equation}
ds^2 = \frac{r^2}{\Rads^2} \,  \eta_{\mu\nu} \, dx^\mu \, dx^\nu 
+ \frac{\Rads^2}{r^2} \, dr^2 + \Rads^2 \, d\Omega_5^2 \ .
\label{e:A5S5}
\end{equation}	
The first two terms here describe the AdS$_5$, a maximally symmetric spacetime of constant negative curvature, with radius\footnote{
Although AdS is spatially non-compact, it has a characteristic size given by its curvature radius:  the Ricci scalar is 
$R=-20 / \Rads^2$. 
Moreover, all timelike geodesics in this spacetime oscillate with period $2 \pi \, \Rads$, so AdS acts like a confining box.
} 
$\Rads$. The last term gives the $S^5$, 
likewise with radius  $\Rads$.
The D-branes are no longer localized within the geometry, but as in \req{e:ebb3}, their effect manifests itself in the 5-form flux through the $S^5$.

The low-energy limit of our D3 brane system is then described by a (10-dimensional) string theory with just closed strings in \adss55\ when $\gst \, N \gg 1$, and by a (4-dimensional) super Yang-Mills $SU(N)$ gauge theory when $\gst \, N \ll 1$.  But since the gauge theory is well-defined at any coupling, it is natural to conjecture that this description in fact applies even when $\gst N$ is large, i.e.\ in the {\it same} regime as where the closed string description holds.

This observation led Maldacena \cite{Maldacena:1997re} to conjecture that
\begin{equation}
{\it 
String \ theory  \ on  \  AdS}_5 \times S^5
\quad \dual \quad
{\cal N} = 4, \ SU(N) \ {\it gauge \ theory \ in \ 4D} \ .
\label{e:ACcorr}
\end{equation}	
Here the `$\dual$' sign indicates a full duality:  the two sides are simply  different languages which describe the {\it same} physics. 

The statement \req{e:ACcorr} quickly became known as the \AC\ correspondence, because AdS specifies the sector of spacetimes for the LHS, and the 4-dimensional gauge theory on the RHS is a conformal field theory (CFT).  In fact, \cite{Maldacena:1997re} gave several other examples which we will mention in \sect{s:generalizations}, all having the \AC\ structure: string or M-theory on AdS$_{d+1}$ (times a compact manifold) is dual to a $d$-dimensional CFT.
Later people started referring to \AC\ alternately as \GG\ (or sometimes gauge/string) duality to emphasize the emergence of gravity (or string theory) from the gauge theory and  to indicate its greater generality.\footnote{
As anticipated already by \cite{Maldacena:1997re}, one can modify the boundary conditions away from AdS and correspondingly the field theory need not be a CFT.   (Also note that in the latter terminology the order of the two sides has been swapped:  ``gravity" or ``string" refers to the AdS side and  ``gauge"  to the CFT side.)
}

We will discuss the meaning of this statement in greater depth in  \sect{s:ACdictionary} below; for now we make a few remarks to qualify the correspondence \req{e:ACcorr} a bit more precisely.  First of all, since the LHS still has dynamical gravity, by ``\STY\ on \adss55" we mean \STY\ on a 10-dimensional spacetime which is {\it asymptotically} \adss55.  In particular, all physically sensible gravitational processes are included on the LHS; for instance one can collapse a black hole in AdS (which will provide an important class of examples that we will revisit below).

Secondly, along with the statement \req{e:ACcorr}, the \AC\ correspondence also specifies how the parameters on the two sides  relate to each other.
On the \STY\ side we have two dimensionless parameters, the string coupling $\gst$ and the curvature scale (in string units) of the spacetime on which the theory lives,  $\Rads/\lst$.  On the gauge theory side we have the rank of the gauge group $N$ and the Yang-Mills coupling $\gYM$, which is more naturally expressed in terms of the `t Hooft coupling\footnote{
For $SU(N)$ gauge theories,  `t Hooft \cite{tHooft:1973jz} showed that there is a smooth limit (known as the `t Hooft limit) which takes $N\to \infty$ while keeping $\lambda \equiv  \gYM^2 \, N$ fixed; in this limit, only the planar Feynman diagrams contribute and the gauge theory becomes effectively classical when recast in certain other variables (namely, it can be described as a classical string theory).}
 $\lambda \equiv  \gYM^2 \, N$.
The relation between these is given by 
\begin{equation}
4 \pi \, \gst = \gYM^2  \sim \frac{\lambda}{N}
\andeq
\frac{\Rads}{\lst} 
= \left( 4 \pi \, \gst \, N \right)^{\! 1/4} \sim \lambda^{1/4}
\label{e:pars}
\end{equation}	
In order to trust the gravity solution (i.e.\ to suppress stringy corrections of the geometry) we need to keep $\Rads$ large in string units, which translates to $\lambda \gg 1$.  
On the other hand, to suppress the quantum corrections, we need to keep $\gst $ small.  
Hence classical gravity is valid in the $N \gg \lambda \gg 1$ regime of the parameter space.  The most conservative version of the duality posits \req{e:ACcorr} in this regime, though it is now widely believed that the equivalence holds even at finite $N$ and $\lambda$ (although direct confirmation is stymied by lack of tools to study the string theory much  beyond perturbations in $1/N$ and $1/\lambda$).

To summarize, the key features of the \AC\ duality are the following.
\begin{itemize}
\item 
There is a mapping between a quantum theory of gravity (namely string theory) and an ordinary (non-gravitational) quantum field theory.  Moreover, the dual description of gravity is manifestly background-independent, except for the AdS boundary conditions.
This should help us solve many long-standing questions in quantum gravity by recasting them in a non-gravitational language.
\item 
\AC\ is a {\it strong/weak coupling duality}: when the gauge theory is strongly-coupled (and hence all perturbative techniques fail), we can study it using the weakly-coupled string theory context.  This has been the main point of \cite{Maldacena:1997re} (and subsequently the most utilized aspect of the correspondence).
\item
The \AC\ mapping is {\it holographic}.  Indeed, the correspondence  provides the most concrete example of the holographic principle, suggested few years earlier by `t Hooft \cite{tHooft:1993gx} and Susskind \cite{Susskind:1994vu}.\footnote{
This bold statement, that in a theory of gravity, we can describe physics in a given region by a theory living on its boundary, was inspired by Bekenstein's entropy bound \cite{Bekenstein:1993dz} based on black hole thermodynamics: since black hole entropy scales with horizon area, the amount of information which can be packed in a spherical region should be bounded by its surface area rather than its volume.
}
Albeit partly inspired by related ideas of \cite{Polyakov:1997tj} (see also \cite{Polyakov:1998ju}),   \cite{Maldacena:1997re} did not emphasize holographic nature of the correspondence (in fact the term `holographic' does not appear in the paper); however, it was noted soon thereafter \cite{Witten:1998qj,Susskind:1998dq}.

\end{itemize}

Though many of the clues hinting at \AC\ correspondence \req{e:ACcorr} had existed before, Maldacena's conjecture \cite{Maldacena:1997re} had taken most by surprise.  Indeed, Juan Maldacena, then a young faculty at Harvard one year after graduating from Princeton, instantly became somewhat of a celebrity in the field. 
The general excitement was evident the following summer at the Strings `98 conference, not only during the talks, but also at the conference dinner when all the participants danced  the `Maldacena'.\footnote{
With the lyrics composed by Jeff Harvey who led the dance, this was a take-off on the popular Spanish dance called the Macarena, producing a rather comical effect which made the occasion doubly memorable.}

The statement of the \AC\ duality is so remarkable that one might well wonder whether there is some loophole in this argument:  could something have gone wrong along the way?  For instance, might there be some sharp transition under varying the coupling $\gst N$ or some non-perturbative effect spoiling the extrapolation?  Albeit unexpected, such effects have been vigorously searched for.  So far, however, the \AC\ correspondence has withstood all tests, and whenever exact calculations are possible on both sides, such as in the maximally supersymmetric case where we can use the tools of integrability \cite{Beisert:2010jr}, a precise match is found.  Nevertheless, the search continues as we develop new tools to understand increasingly more general context.

Another (more distanced and vague) level at which the \AC\ correspondence has been questioned by its critics refers to the foundations themselves:  \AC\ has been derived within \STY, but what if \STY\ itself is incomplete or incorrect?  Rather than delving into an involved discussion attempting to dispel such objections here, we will now argue that the \GG\ duality actually stands on its own in a self-contained manner,  independently of \STY.

\subsection{More modern perspective}
\label{s:ACmodern}

Maldacena's derivation notwithstanding, many of us experience some bewilderment at our first encounter of the \AC\ duality.
Indeed, the assertion that a higher dimensional gravitational theory could be fully described by a  lower-dimensional non-gravitational one seems quite preposterous.  
Moreover, it was generally felt in the \STY\ community that we understand quantum field theories well, and that they cannot be so well disguised as  to actually behave like quantum gravity.
 In fact, much of the initial effort in this field stemmed from attempts to find a clear `counter-example' which would indicate that the two sides of \req{e:ACcorr} cannot possibly be equivalent.  Though all such attempts of course failed, this was a tremendously useful exercise for developing our understanding of how the correspondence works.
Before describing some of these checks and entries in the AdS/CFT dictionary, let us pause to see why the correspondence is not obviously wrong, by mentioning a more direct route toward the duality, one which is independent of string theory.

In hindsight, a plausible way to arrive at the \GG\ duality is the following:\footnote{
The presentation here is  largely based on the excellent review of the \GG\ duality by Horowitz \& Polchinski \cite{Horowitz:2006ct}.}
The prospect that a gauge theory might capture a gravitational theory is partly inspired by the matter content: a natural guess, bolstered by representation theory, is that the spin-two graviton could arise as a composite of two spin-one gauge bosons.  Of course, this would be precluded by the Weinberg-Witten no-go theorem \cite{Weinberg:1980kq}, if the graviton and gauge bosons lived in the same Lorentzian spacetime.  However, the holographic principle suggests that the graviton may naturally propagate in higher dimensions, thereby eluding the no-go argument.  The gauge theory would then have to contain not just the graviton but also an extra dimension, namely some quantity with respect to which the physics behaves locally.  One such quantity in a gauge theory is the energy scale:  the renormalization group equation governing the flow of the coupling constants with energy scale is a nonlinear differential equation which is local in the energy scale.  

Having identified a gauge theory quantity which could describe an extra direction in the bulk dual, we want to further ensure that this new `dimension' can be macroscopic.  Since perturbative gauge theory looks nothing like classical gravity, one is led to suspect that the gauge theory should be highly quantum (i.e.\ strongly coupled) in order to reproduce gravity.\footnote{
Anticipating the string theory context, strong coupling can also be motivated by the requirement of keeping only spin-two graviton while removing all higher-spin composite objects.
}
Taking this guess seriously, one would then seek a gauge theory where the coupling remains strong over a large range of energies, in order to recover an extra dimension of macroscopic size.  This is most easily achieved in a conformal field theory where the coupling does not run, allowing for an infinite holographic direction.  
Nonetheless, the possibility of having excitations with various energies propagating in higher dimensions still seems to have more degrees of freedom than can be accommodated within a lower-dimensional theory.  In order to overcome this obstacle, the gauge theory then has to contain sufficiently large number of degrees of freedom, which one may hope to  achieve with a large gauge group;  for $SU(N)$ gauge theories, this suggests the  't Hooft limit \cite{tHooft:1973jz}, namely taking $N \to \infty$ keeping $\lambda \equiv \gYM^2 \, N$ fixed (but large by the previous argument).

To keep our strongly-coupled gauge theory under control, it is convenient to impose supersymmetry (since this keeps the Hamiltonian bounded from below and thereby precludes many potential instabilities).  The most supersymmetric case is ${\cal N}=4$ (which has 4 copies of the minimal 4-dimensional supersymmetry algebra), which simultaneously ensures conformal invariance:  The 4-dimensional  ${\cal N}=4$ $SU(N)$ gauge theory, \SYM,\footnote{
This theory can be obtained from a dimensional reduction of a 10-dimensional $SU(N)$ gauge theory with 16 supersymmetries with smallest algebra, 
$ {\cal L} = \frac{1}{2\gYM^2}  {\rm Tr} ( F_{\mu\nu} \, F^{\mu\nu}) + i \, {\rm Tr} ( {\bar \psi} \, \gamma^\mu \, D_\mu \psi )$ (with both the gauge field $A_\mu$ and the spinor $\psi$  in the adjoint $N\times N$ representation).  Upon dimensional reduction to 4 dimensions, the $A_\mu$ decomposes into a 4-dimensional gauge field and 6 scalars $\ph^i$ which are symmetric under $SO(6)$ rotation, while the spinor $\psi$ separates into four 4-dimensional Weyl spinors.
This 4-dimensional SYM is not only a finite theory, but indeed the simplest interacting 4-dimensional field theory.
} is a conformal field theory (CFT).  When formulated on Minkowski space, 
$ds_{\scriptscriptstyle {\rm CFT}}^2 = \eta_{\mu\nu} \, dx^\mu \, dx^\nu$,
the vacuum is invariant under \Poinc\ transformations, and by virtue of conformal invariance it is also in particular invariant under a rigid scale transformation $x^\mu \to \alpha \, x^\mu$, which simultaneously rescales the energy $E \to E/\alpha$.  Identifying inverse energy with the extra dimension (labeled by a new coordinate $z$), the most general 5-dimensional bulk metric consistent with these symmetries is AdS$_5$:
\begin{equation}
ds^2 = \frac{\Rads^2}{z^2} \, \left(  \eta_{\mu\nu} \, dx^\mu \, dx^\nu + dz^2 \right)
\label{e:AdSPoinc}
\end{equation}	
where we've rescaled $z \sim 1/E$  so as to express the metric in terms of one free parameter, the AdS scale $\Rads$.  A trivial change of variables, $z=\Rads^2/r$, recasts \req{e:AdSPoinc} into the  form used in \req{e:A5S5}; i.e.\ $r$ corresponds to energy scale as indeed identified by \cite{Maldacena:1997re}.

So far, the above arguments motivate the equivalence of gravity on 5-dimensional AdS with 4-dimensional  ${\cal N}=4$ $SU(N)$ gauge theory, along with the expectation that the number of `colors' $N$ should be related to the size of AdS, $\Rads$.  But one can go even further.   
To match the supersymmetry of the gauge theory side, we need to extend gravity to the full supergravity, which is naturally formulated in 10-dimensions and has a solution \adss55.  So we get not just one, but six, extra dimensions, five of which (making up the $S^5$) are naturally associated with the scalars $\ph^i$ of the 4-dimensional \SYM.
Moreover, if we probe further still, by considering highly boosted bulk states, the corresponding excitations on the \GT\ side exhibit a one-dimensional structure which  can be associated with excitations of a string \cite{Berenstein:2002jq}.
 So by making the full structure self-consistent, one could in principle `uncover' string theory.\footnote{
 In fact, the idea of strings arising from large-$N$ (planar) limit of gauge theory  was already suggested by 't Hooft \cite{tHooft:1973jz}; the \AC\ correspondence renders this occurrence fully explicit.
}  Hence in hindsight, the \AC\ duality \req{e:ACcorr} emerges quite naturally.
Moreover, albeit puzzling, it elegantly explains why apparently different results in gauge theory and gravity happen to give the same answers.  In a sense, having so many coincidences with no explanation would have been far more perplexing...

Finally, at a broader level, it can be argued \cite{Marolf:2008mf} that quantum gravity with asymptotically AdS (or even asymptotically flat) boundary conditions is holographic, in the following sense: due to diffeomorphism invariance, the gravitational Hamiltonian is a purely boundary term, expressed as a surface integral at infinity.  Since it is the operator which generates time translations, any set of boundary observables which are available at any given time are necessarily available on the boundary at any other time.  More technically, Marolf  \cite{Marolf:2008mf} shows that (at least at the perturbative level) there is a complete algebra of boundary observables within any neighborhood of any boundary Cauchy surface.
This idea is referred to as boundary unitarity, and has important implications for example about the black hole information paradox.

\subsection{Early checks and entries in the \AC\ dictionary}
\label{s:ACdictionary}

Now that we have explained two separate arguments for the \GG\ duality \req{e:ACcorr}, let us turn to its implications.  Of course, even if rigorously proven, the correspondence would be of little use if we didn't know how to relate physically interesting quantities between the two sides.  Much the subsequent (and ongoing) effort went into establishing the AdS/CFT dictionary. 
The basic entries in this dictionary follow immediately from the above arguments.  Let us therefore start by specifying these more explicitly.

\paragraph{Symmetries:}  
Perhaps the most immediate check is that the symmetries match between the two sides.
Let us start with \adss55.  The isometry group of AdS$_5$ is $SO(4,2)$, which is manifest when we write AdS as the embedded hyperboloid
\begin{equation}
-X_{-1}^2 - X_{0}^2 + X_{1}^2 + 
 \ldots  
+ X_{4}^2= - \Rads^2
\label{}
\end{equation}	
in $\RR^{4,2}$ with metric
$ds^2 = -dX_{-1}^2 - dX_{0}^2 + dX_{1}^2 + \ldots + dX_{4}^2 $; and similarly, the isometry group of the $S^5$ is $SO(6)$, so the full
(bosonic) symmetry is $SO(4,2) \times SO(6)$.
From the CFT side, the conformal group in four dimensions is $SO(4, 2)$ (which includes \Poinc\ transformations as well as scale transformations and special conformal transfomations), and the six scalar fields $\ph^i$ and four fermions are related via a global $SU(4) \simeq SO(6)$ R-symmetry.  Both sides also have 32 supersymmetries, which manifest themselves as Killing spinors in  \adss55\ on the gravity side, and as superconformal algebra on the \GT\ side.

\paragraph{Spacetime directions:}  
The above symmetry matching immediately suggests how the 10 dimensions of the bulk \adss55\ map to the CFT.  From the $SO(6)$ symmetry we see that the $S^5$ directions  are captured by the scalar fields $\ph^i$.   The radial bulk direction $r = \Rads^2/z$ we have already  associated with the energy scale in the \GT.  
Therefore the boundary of AdS, $r=\infty$ or $z=0$, is naturally identified with the UV of the \GT.  We will elaborate on this relation further below.
What remains are the transverse directions $x^\mu$ of AdS, but these map directly to the Minkowski spacetime dimensions the \GT\ lives on.
In particular, the time $x^0$ on both sides gets naturally identified, which means that the notion of Hamiltonian should likewise match on both sides.

So far, we have indicated how the mapping between bulk spacetime directions and \GT\ quantities   works for pure \adss55.   Even for this highly symmetric case, however,  it is far from obvious that the relevant \GT\ quantities should be related to each other so as to uphold bulk diffeomorphisms.  In other words, to a bulk observer in \adss55, all 9 spatial directions look locally the same, while in the \GT\ this local symmetry is totally obscure.  
This becomes much more poignant when the geometry breaks all the symmetries and only asymptotes to \adss55: in such generic case the bulk-boundary map is no longer easy to specify. 
For example, even though there is a natural notion of `constant time' slice on the boundary (up to overall boosts), for generic  time-dependent bulk geometry there is no correspondingly uniquely defined time-foliation in the bulk.  
Moreover, while the bulk spacetime is dynamical, governed by 10-dimensional Einstein's equations with appropriate matter content, the boundary spacetime is fixed: there is no sense of gravitational backreaction in the \GT; so in this sense even the $x^\mu$ directions `emerge' nontrivially.
Indeed, understanding how bulk locality emerges from the \GT\ is one of the important open questions in this field, which would teach us something nontrivial about the \GT\ as well as about the nature of spacetime.

The AdS metric specified by the $(x^\mu,z)$ coordinates in \req{e:AdSPoinc}, commonly known as the \Poinc\ (patch of) AdS, is geodesically incomplete.   The \Poinc\ horizon, $z \to \infty$, is a regular null surface, and we can extend the spacetime to its global AdS form; in static, spherically symmetric coordinates this can be written as
\begin{equation}
ds^2 = - \left( \frac{\rho^2}{\Rads^2} + 1 \right) \, d\tau^2 + \frac{d\rho^2}{\left( \frac{\rho^2}{\Rads^2} + 1 \right) }
+ \rho^2 \, d\Omega_3^2
\label{e:gAdS}
\end{equation}	
Although both \Poinc\ AdS \req{e:AdSPoinc} and global AdS \req{e:gAdS} are explicitly static, the timelike Killing fields $\dda{t}$ and $\dda{\tau}$ are distinct.  
\begin{figure}
\begin{center}
\includegraphics[width=5.5in]{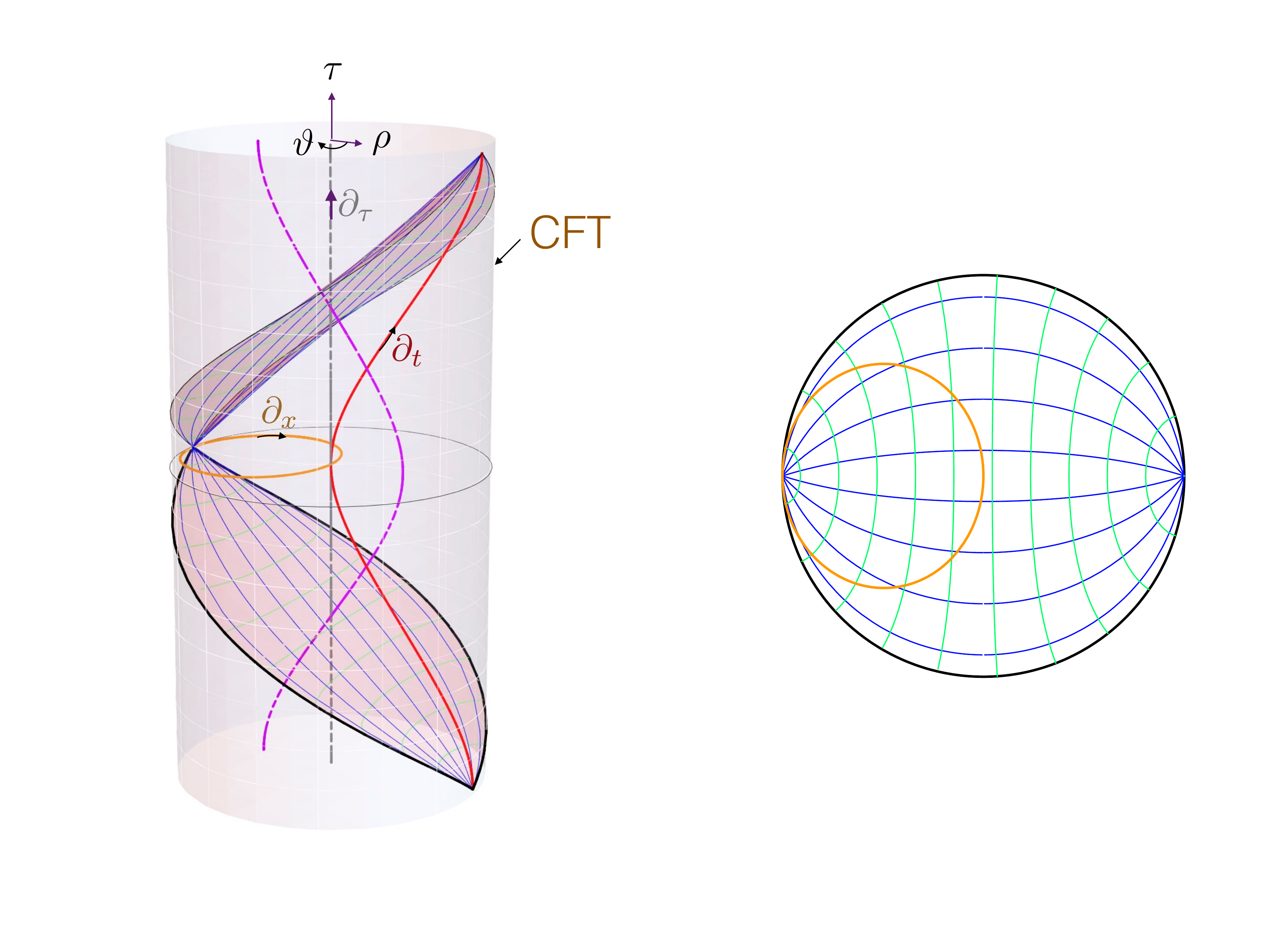}
\caption{
{\bf Left:}
Schematic plot of global AdS$_5$, showing the \Poinc\ patch.  The vertical coordinate is $\tau$, the horizontal radial coordinate is $\tan^{-1} \rho$ (so that the spacetime is drawn compactified and radial null geodesics are at 45 degree angle), and the angular coordinate $\vartheta$ is one of the $S^3$ directions of the AdS.
Poincare horizon is a null surface [red] (generated by null geodesics [blue]) whose constant-$\tau$ slices are spacelike extremal surfaces anchored on spherical regions or (on equatorial slices of $S^3$) spacelike geodesics [green].
For orientation we have also plotted two $z = \Rads$ (equivalently $r = \Rads$) curves, one at constant $x^i=0$ which describes orbit of $\dda{t}$ [red $\partial_t$], and one at constant $t=0$ which describes orbit of $\dda{x}$ [orange $\partial_x$] Killing fields. 
Also shown are two timelike geodesics one static at `center' of global AdS following $\dda{\tau}$ [grey $\partial_\tau$] and one [purple] reaching to $\rho = \Rads$, to illustrate the confining nature of AdS. 
{\bf Right:}
Poincare disk of AdS, which is a spatial (constant $\tau$) slice of AdS. Spacelike geodesics and projections of null geodesics, as well as the $z=\Rads$, $t=0$ curve, are drawn [with the same color scheme as in the left panel].
}
\label{f:AdS}
\end{center}
\end{figure}
This is indicated in \fig{f:AdS} which presents a sketch of global AdS, showing a \Poinc\ patch, several geodesics, and an indication of the corresponding coordinates.

We can take \req{e:ACcorr} to apply to string theory on the full global \adss55, in which case the \GT\ is formulated on the Einstein Static Universe $S^3 \times R$.
Notice that in both cases, \Poinc\ and global, the \GT\ can be naturally thought of as `living on the boundary' of AdS:  it is formulated on a spacetime which is in the same conformal class (the relevant structure for a conformal field theory) as the boundary metric induced from the bulk.  For example, multiplying the line element \req{e:gAdS} by $\Rads^2 / \rho^2$ and taking $\rho \to \infty$ with $d\rho = 0$ yields the Einstein static universe $ds_{\scriptscriptstyle {\rm CFT}}^2 = - d\tau^2 + \Rads^2 \, d\Omega_3^2$.
Moreover, the fact that the conformal compactification of AdS has a timelike boundary allows for a Lorentzian field theory on this background.
We will continue using  terminology  `CFT living on the boundary of AdS' as a conceptual crutch, since it will render many of the other elements of the \AC\ dictionary easier to understand intuitively.  We should however keep in mind that the \GT\ encodes the entire bulk, not just what is happening asymptotically.

\paragraph{Scale/radius duality:}  
Closely-related to the above remarks is the statement of scale/radius duality, also known as the UV/IR duality.  
Recall that motion in the radial direction $r$  in \req{e:A5S5} corresponds to moving in energy scale in the dual field theory.  For pure AdS, one can see this at the level of the symmetry, as motivated in \sect{s:ACmodern}: high energies (or short distance, i.e.\ UV) on the boundary is associated with large radius (so in this sense IR) in the bulk. In particular, a UV cutoff in the boundary corresponds to an IR cutoff in the bulk.  
This correspondence was put on a firmer footing in \cite{Susskind:1998dq}, which showed that when suitably regulated, the \GT\ provides a  holographic description with one bit of information per Planck area.
Operationally, if instead of energy scales on the boundary one thinks of length scales, in the asymptotic region one can associate the length between endpoints of a given spacelike geodesic with the bulk radial position to which this geodesic penetrates (c.f.\ the green curves on the right panel of \fig{f:AdS}).

The scale/radius duality provides a good guide to our expectations of various physical processes.  For example, a bulk particle which falls deeper into AdS (due to the attractive potential caused by the AdS curvature) is described by a localized excitation in the \GT\ which spreads out with time.  In the presence of a black hole, the bulk particle nearing the horizon can be regarded as the \GT\ excitation thermalizing \cite{Banks:1998dd}.
However, the precise form of the scale/radius duality fuzzes out for anything other than pure \Poinc-AdS geometry, so for a generic bulk spacetime it is most useful only asymptotically; deeper in the bulk or for rapidly-evolving processes it need not provide a reliable guide, as indicated below.

\paragraph{Causality:}  
Convenient as the scale/radius duality is for conceptual picture of part of the bulk-boundary mapping, too-naive applications of it  can however lead to apparent contradictions, as instructively exemplified by \cite{Horowitz:1999gf}, one of the earliest explorations of \AC\ in manifestly Lorentzian context.
The authors studied how the CFT manages to reproduce bulk causality, by considering gedanken-experiments involving massless particles in the bulk:  Suppose we send two particles radially inward from the same boundary position but at different times.  The second particle is always to the future of the first one and therefore cannot influence its dynamics.  On the other hand, in the CFT where one expects each particle to be described by  some initially localized disturbance which spreads out (and hence the disturbance produced by the second particle to overlap with that of the first one), it appears quite plausible that second can influence the first.  If this happened, the CFT would violate bulk causality.   One can also imagine sending two particles (non-radially inward) from different positions on the boundary but at the same time.  In the CFT, the two disturbances interact on a time scale given by their initial separation, whereas in the bulk, if the particles miss each other, one would  expect them to interact on a much longer time scale, if at all.

Both of these puzzles get resolved by correctly accounting for the gravitational backreaction of the particles, upon which the AdS and CFT answers match exactly.  In particular, \cite{Horowitz:1999gf} shows that a massless particle produces a gravitational shock wave, which in the CFT
gives a `light-cone state', a disturbance which is localized on light cone of the boundary spacetime.  
So in the first example, the two CFT excitations indeed do not interact, whereas in the second example the shock waves produced by the particles do interact on the same timescale as predicted by the CFT.
The lesson from this resolution typifies many of the others: once the physics on both sides is understood correctly, we indeed get a perfect agreement.  As a bonus, a physical effect which may be quite subtle on one side is often completely obvious on the other side of the correspondence.   In this way, the recasting of a given setup in the dual picture usually elucidates the physics.

Coming back to the specific issue of causality, one might nevertheless wonder how can bulk causality generically coincide with CFT causality, in spite of  the bulk having more directions.
Since the boundary metric (which determines the causal relations in the CFT) is induced from the bulk metric, it is clear that any two events which are causally related along the boundary must likewise be causally related in the bulk.  However the converse is far less obvious:  could one not travel through the bulk faster than around the boundary (as would be the case for a cylinder in flat spacetime), thereby violating CFT causality?  It is easy to check that this is not the case for pure global AdS: as evident from \fig{f:AdS} (blue curves), all null geodesics from a given boundary point, whether passing through the bulk or along the boundary,  in fact reconverge at the same boundary point (namely antipodally on the sphere and time $\pi \, \Rads$ after the starting point). 
Given that in pure AdS it takes the same time for a light ray to go through the bulk as along the boundary, one might worry that a small deformation to the bulk geometry could then speed up the bulk geodesic and therefore send signals outside of the CFT light cone.
That this does not happen was proved by Gao \& Wald \cite{Gao:2000ga} who showed that physically sensible\footnote{
More precisely, \cite{Gao:2000ga} assume the bulk satisfies the null energy condition (and the null generic condition).
} deformations of AdS would lead to a gravitational time delay (as opposed to time advance) of bulk geodesics, thereby ensuring that causal processes in the bulk cannot violate CFT causality.

\paragraph{Observables and correlation functions:}  
Now that we have seen how the bulk spacetime as such relates to the  \GT, what about various bulk fields propagating on this spacetime?
At the most basic level, the answer is quite simple: every bulk field $\phi$ corresponds to an operator ${\cal O}$ in the \GT.
This allows us to extract a useful entry in the \AC\ dictionary, namely how the observables of the \GT\ are encoded in the bulk.  
This has been worked out in the seminal papers  \cite{Gubser:1998bc,Witten:1998qj} within few months after \cite{Maldacena:1997re} appeared, and the one-to-one identification between the two sides developed therein
provided important early checks of the \AC\ correspondence.

One natural set of observables is given by expectation values of local gauge invariant operators (constructed as single trace of a local product of \SYM\ fields); they can be generally associated with string states, but a certain subset of these correspond to supergravity fields in the bulk.  More specifically, since local operators are associated with the UV of the theory, it is not surprising that expectation values of such operators correspond to the asymptotic behavior of the corresponding bulk fields.  In particular, a normalizable bulk field which falls off as $\phi \sim \mu/r^\Delta$ corresponds to an operator ${\cal O}$ with scaling dimension
$\Delta$ (which depends on the mass and type of the field and spacetime dimension), with expectation value $\vev{{\cal O}} = \mu$.
The rate of the bulk field fall off is related to the nature on the field theory perturbation.\footnote{
More specifically, irrelevant (in the UV) perturbations of the field theory correspond to massive modes in supergravity (which fall off quickly), marginal perturbations correspond to massless modes, and relevant perturbations correspond to modes with negative mass squared; unlike for asymptotically flat spacetime, the latter is allowed in a certain range of masses (above the so-called Breitenlohner-Freedman bound) \cite{Breitenlohner:1982bm,Breitenlohner:1982jf} without producing any instability.}

To substantiate this and obtain more general correlation functions, imagine adding a `source' term to the CFT Lagrangian,
$\int d^4 x \, \phi_0(x) \, {\cal O}(x)$. 
Its exponential gives a generating function of correlation functions (i.e.\ $\vev{{\cal O}(x) \cdots {\cal O}(y)}$ is obtained by taking functional derivatives with respect to $\phi_0$'s and then setting $\phi_0=0$),
which one can naturally associate with the partition function of the string theory, as a functional of the boundary condition $\phi_0$,
\begin{equation}
{\cal Z}_{\scriptscriptstyle {\rm string}}[\phi_0] = \vev{ e^{\int d^4x\, \phi_0\, {\cal O}}}
\label{corgen}
\end{equation}	
In effect, one can rephrase the \AC\ duality in terms of equivalence between the field theory partition function (viewed as function of sources for each operator) and quantum gravity partition function (viewed as a function of boundary conditions for each bulk field).
In the classical limit one can approximate the LHS by the exponential of the classical action for the solution specified by the boundary condition $\phi_0$.
One can then think of correlation functions in the \GT\ between operators ${\cal O}$ inserted at certain points on the boundary as propagation of corresponding fields in the bulk between these points.

To determine which boundary operators map to which bulk fields, one can use the symmetries.  The two objects must have the same Lorentz structure and quantum numbers.  For example, conserved currents in the \GT\ are associated to global symmetries, so the corresponding sources act as external background gauge fields, which are boundary values of a dynamical gauge field in the bulk.  From the gravitational standpoint, the most crucial example of this correspondence is that bulk gravitons $h_{\mu\nu}$ naturally couple to boundary stress energy momentum tensor $T^{\mu\nu}$.  One can think of this as a generalization of the familiar fact that the ADM mass in asymptotically flat spacetime can be extracted from the leading radial fall-off in the metric.  (For a more detailed review, see e.g.\ \cite{Fischetti:2012rd}.)

One might naively expect that just knowing the asymptotic values of bulk fields does not tell us much about their structure in the bulk; but it turns out that the situation is in fact much better in \adss55\ than it would be in flat spacetime, because AdS effectively acts like a lens which refocuses some information to be extractible from just the leading fall-offs.\footnote{
For example, using operators involving $S^5$ spherical harmonics, one can read off the size of a small  (compared to $\Rads$) spherical object from local field theory expectation values
\cite{Horowitz:2000fm}.
This may seem surprising in light of the scale/radius duality which would naively suggest that any information about sub-AdS scales would require highly delocalized observables in the \GT, but the operators we use for this task have very high dimension, which gives large dispersion for an actual `measurement' \cite{Hubeny:2000eu}.
}
Nevertheless, we can gain far more information about the bulk physics from less local observables, such as the above-mentioned correlation functions.  Roughly-speaking, the further the separation between the insertion points on the boundary, the greater its sensitivity to physics deeper in the bulk, as might indeed be anticipated from the scale/radius duality.  For 2-point function of high-dimension operators, one may use a WKB approximation to express the bulk Green's functions in terms of geodesics; as clear from \fig{f:AdS} (green curves), these typically penetrate deeper for greater separation of endpoints.
We will mention an interesting application of this in \sect{s:BHs}.

Another important class of nonlocal observables in the \GT\ are  Wilson loops, $W({\cal C})$, specified by some closed loop ${\cal C}$ on the boundary.  Their expectation values for example allow us to compute the quark anti-quark potential (with the quark and anti-quark trajectories given by ${\cal C}$).
A Wilson loop is defined by a path-ordered integral along ${\cal C}$ of a gauge connection $A_{\mu}$, schematically\footnote{
In the \AC\ context, this needs to be generalized  by also including the \GT\ scalars \cite{Maldacena:1998im}. 
}
$W({\cal C}) = {\rm Tr} \left[ P \, \exp \left( i \oint_{\cal C} A \right) \right]
$
with the trace taken over some representation of the gauge group.  
Though this seems like quite a complicated object in the field theory, it turns out that its bulk dual is remarkably simple:
For the fundamental representation, the leading order contribution comes from the proper area of a string worldsheet in the bulk \cite{Maldacena:1998im}, which describes a 2-dimensional\footnote{
In fact, certain other representations can be better characterized using D3-branes, and hence by corresponding 3-dimensional extremal surfaces \cite{Drukker:2005kx} (though in even more exotic cases, with sufficiently many such branes, the bulk is better described by regular but topologically non-trivial `bubbling' geometries \cite{Lunin:2006xr}).
} extremal surface anchored on ${\cal C}$.  
Ultimately this correspondence is rooted in the deep connection between a \GT\ flux tube and a bulk string, which we will revisit in \sect{s:appliedAC}.
For simple enough cases one can compute $\vev{W({\cal C})}$ exactly in the \GT, and confirm precise agreement with the bulk string worldsheet area calculation \cite{Drukker:2000rr}.

The above examples demonstrate that interesting field theory observables can actually be obtained from elementary geometrical bulk constructs,  such as areas of extremal surfaces.  This has proved invaluable in computing these field theoretic quantities, since the dual calculation is typically much easier than attempting a more direct approach.  Conversely, if one were given this  `data' in a field theory with a holographic dual, one could in principle use it to probe the corresponding bulk geometry.  

\paragraph{Entanglement entropy:}  
Before proceeding with the early checks of \AC, we pause briefly to mention a more recently-considered (and rather different type of) quantity in the field theory, which is nevertheless thematically related to our story by having a simple geometric dual description: the entanglement entropy. Entanglement is arguably the most non-classical manifestation of quantum mechanics, 
which can be used as a resource for performing tasks that cannot be accomplished with classical resources, such as quantum teleportation.  It is actually used in a wide range of subjects, including quantum information theory, quantum optics, condensed matter physics, etc..  
A particularly convenient measure of entanglement is {\it entanglement entropy}, specified by a subsystem ${\cal A}$ and a total state $\rho$ of the system.  Formally, it is defined as the Von Neumann entropy $S_{{\cal A}} = - {\rm Tr} \left( \rho_{{\cal A}} \log \rho_{{\cal A}}\right)$ of the reduced density matrix $\rho_{{\cal A}}$ obtained by tracing $\rho$ over the complement of ${\cal A}$.  
In a local field theory, the subsystem ${\cal A}$ can be specified by a given spatial region, bounded by an `entangling surface' $\partial {\cal A}$.  Entanglement entropy of ${\cal A}$ then contains information  about the spatial distribution of quantum correlations in the system.

In all but the simplest systems, \EE\ is however extremely difficult to compute, and even harder (if not impossible) to measure, being sensitive to detailed correlations in the state.  Remarkably, here too \AC\ comes to the rescue.  It was proposed by Ryu \& Takayanagi \cite{Ryu:2006bv}
that in static situations, the entanglement entropy $S_{\cal A}$ is given by quarter of the area (in Planck units) of a certain co-dimension 2 bulk surface, analogously to the black hole entropy.  The surface in question is a minimal surface at constant time which is anchored on the entangling surface $\partial {\cal A}$.  This proposal, recently proved \cite{Lewkowycz:2013nqa} using Euclidean path integral techniques, has 
rendered many non-trivial statements in the field theory (such as the strong subadditivity property of entanglement entropy \cite{Headrick:2007km}) beautifully manifest in the geometrical language.  
Since entanglement entropy is a well-defined quantity even in time-evolving situations (corresponding to time-dependent bulk geometries), the construction of \cite{Ryu:2006bv} is unnecessarily restrictive.  To overcome this limitation, \cite{Hubeny:2007xt} generalized the proposal to arbitrary asymptotically AdS bulk geometries, using the guidance of general covariance: the \EE\ is given by the (quarter-)area of a bulk co-dimension 2 {\it extremal} surface anchored on 
$\partial {\cal A}$.\footnote{
There are two additional specifications which give the holographic \EE\  a more non-local flavor:
1) In case of multiple extremal surfaces anchored on $\partial {\cal A}$ (which can easily happen in e.g.\ black hole geometries), the requisite surface is the one with the least area. 
2) More intriguingly, the extremal surface should be homologous to the specified region ${\cal A}$ (though this property needs further qualifications, as illustrated in \cite{Hubeny:2013gta}).
While harder to prove, this covariant prescription has however also passed many non-trivial checks; for example, it implies strong subadditivity \cite{Wall:2012uf} and it is non-trivially consistent with field theory causality \cite{Headrick:2014cta}. }

There seems to be something quite deep and mysterious about this relation between quantum entanglement on the one hand, and classical geometry on the other, which we will revisit in \sect{s:appliedAC}.  
At a more pragmatic level, we note that
as in the case of more conventional CFT observables, \EE\ allows us to probe the bulk geometry.  In particular, if we know the areas of extremal surfaces anchored on a family of entangling surfaces, we can in certain circumstances invert this relation to determine the bulk metric.

\paragraph{States and geometries:}  
Having discussed how one can probe the geometrical structure of asymptotically \adss55\ spacetimes using various quantities in  the \GT, let us consider from the \GT\  point of view what {\it gives rise} to the bulk spacetime deformations in the first place.  In the \GT, the only state which respects all the symmetries is the vacuum; hence pure \adss55\ corresponds to the vacuum of the field theory.  If we consider some excited state, the symmetries will be broken, so correspondingly the bulk geometry should be deformed as well.  However, to see this at the classical geometry level, we need the energy of the excitation to scale as $N^2$, since the 10-dimensional Newton's constant in AdS units is $\GN = \gst^2 \, 
\lst^8 \sim \Rads^8 / N^2$.  

Lower energy (parametrically smaller than ${\cal O}(N^2)$) excitations will not backreact on the geometry; instead, they can be described by fields propagating on \adss55.  
But these are the quantities we have already considered above.
As shown by Witten \cite{Witten:1998qj}, 
all linearized supergravity states (perturbations of \adss55) have corresponding states in the gauge theory; for example the Kaluza-Klein modes of the supergravity fields in the bulk are identified with certain simple operators in the CFT, with excitation spectra matching on both sides.
As one increases the energy of the excitations to ${\cal O}(N)$,  one encounters an interesting effect: massless string states with high angular momentum $J$ blow up into spherical D3-branes whose size grows with $J$.\footnote{
This is related to the `Myers effect' \cite{Myers:1999ps} wherein polarized D-branes blow up into a sphere. 
}  This gives rise to  a new class of supersymmetric states, called `giant gravitons', which can wrap inside the internal sphere \cite{McGreevy:2000cw} or the 3-sphere in AdS \cite{Hashimoto:2000zp}.\footnote{
Subsequently, \cite{Corley:2001zk} provided a unified description on the gauge theory side in terms of gauge invariant operators, conveniently characterized by Young diagrams.}
They are stabilized by the angular momentum, but in the former case  there is a maximal angular momentum $J \sim N$ for which they can still `fit' into the spacetime.
These states were shown to be in one-to-one correspondence with classical supersymmetric solutions in the \SYM\ carrying the same quantum numbers, but additionally they provided an explanation of the previously mysterious stringy exclusion principle \cite{Maldacena:1998bw} which had been derived from the CFT side: because the relevant states are constructed from traces of products of $N \times N$ matrices, there can be at most $N$ independent ones.  
Since this match is nonperturbative in $1/N$, it provided a nice check that the duality \req{e:ACcorr} remains valid even at finite $N$.

Let us now return to the higher-energy ($\propto N^2$) \GT\ excitations which do deform the bulk \adss55\ geometry.  In fact, since there are ${\cal O}(N^2)$ degrees of freedom in the \GT, such excitations are rather natural: we can obtain them for instance by exciting each degree of freedom by some small amount.
Although the precise identification of a specific ${\cal O}(N^2)$ excitation in the field theory in terms of the bulk geometry is in general very difficult, one can in fact retain full control over certain  special, but nevertheless rather rich, class of supersymmetric states, describable by free fermions \cite{Berenstein:2004kk,Corley:2001zk}.  These turn out to have an explicit one-to-one mapping to the so-called LLM geometries \cite{Lin:2004nb}, specified by arbitrary closed curves in 2-dimensions (which correspond to the Fermi surface of the fermions, and simultaneously prescribe the boundary conditions which uniquely determine the 10-dimensional bulk geometry).  These geometries are smooth and horizonless, but can have interesting topological structure.

For more general excitations, we usually content with considering various coarse-grained features of their bulk duals.  Such excited states (in the boundary CFT) have a non-zero expectation value of the (boundary) stress tensor $T^{\mu\nu}$, which can be thought of as arising from the bulk metric being deformed away from pure AdS. The stress tensor, then, provides a useful characterization of the bulk geometry.\footnote{
One might hope that it would actually allow us to determine the bulk metric fully, since by using holographic renormalization group ideas, one can write the metric as a radial series expansion around the asymptotic behavior \cite{deHaro:2000xn}, but in general this series may not converge; in fact it generically leads to naked singularities in the bulk \cite{Bhattacharyya:2008jc}.
}  We have already encountered an example, namely the light-cone states of \cite{Horowitz:1999gf} being dual to gravitational shock waves in the bulk, but such states are still rather special.  

The most {\it generic} state with energy of ${\cal O}(N^2)$, which by the 2nd law of thermodynamics will be in thermodynamical equilibrium, in fact corresponds to a bulk black hole.  This is easy to see at the most basic level, as the end state of a generic process: in the \GT\ a generic high energy excitation will thermalize, while in the bulk, the combined effects of backreaction and AdS attractive potential will force a generic excitation to collapse to a black hole.
So, roughly-speaking, AdS black holes correspond to thermal states in the \GT.  This illustrates an important point that although the theories appearing on the two sides of the \AC\ correspondence \req{e:ACcorr} are supersymmetric, the actual  states we consider need not be.  So supersymmetry is {\it not} a necessary ingredient in mapping between the two sides of the duality.
Since apart from the above genericity argument, black holes also provide  a particularly useful arena to consider in the context of both elucidating the \AC\ correspondence (and hence \QG), as well as for the applications discussed in \sect{s:appliedAC}, we will devote a separate subsection to this important topic.  

\subsection{Black holes in AdS}
\label{s:BHs}

We have already evoked (extremal) black branes as a key step in Maldacena's derivation \cite{Maldacena:1997re} of the \AC\ correspondence; but once in the low-energy limit, the black hole was no longer evident:  instead, we were left with a \Poinc\ patch of AdS (with the black brane horizon becoming the \Poinc\ horizon), which we could then globally extend to the causally-trivial global AdS geometry.

The black holes we now wish to consider are {\it within} this asymptotically AdS spacetime, appearing as new objects.\footnote{
The large-mass limit of a spherical black hole wherein the horizon becomes translationally invariant, namely the planar \schw-\adss55\  geometry, can in fact be obtained directly from a D3 brane system, as already described in \cite{Maldacena:1997re}, by suitably exciting the D3 branes and then taking the low energy limit of a near-extremal black 3-brane geometry.  As we will see below, such a black hole is most naturally related to \Poinc-AdS.
}
  We will focus on a neutral, static, spherically symmetric black hole in AdS, i.e.\ the global \schw-\adss55\  solution, which is the easiest and most illustrative example.   
We will start by describing the geometry, thermodynamics, and causal structure, then briefly discuss time-dependent context, and finally revisit the previously-discussed issue of whether/how does the \GT\ encode the region behind the horizon.

\paragraph{Geometry:}  
The \schw-\adss55\  geometry describes a spherical black hole in global AdS, trivially smeared over the $S^5$ in a direct product structure.
The solution can be obtained either as a 10-dimensional solution to Einstein's equations with self-dual 5-form field strength with flux given by $\Rads$, or, by reducing on the $S^5$, as a solution to 5-dimensional Einstein's equations with negative cosmological constant $\Lambda = - \frac{6}{\Rads^2}$ but zero bulk stress tensor.
The resulting metric is given by
\begin{equation}
ds^2 = - g(r) \, dt^2 + \frac{dr^2}{g(r)} + r^2 \, d\Omega_3^2 + \Rads^2 \, d\Omega_5^2
 \ \ , \qquad
g(r) = \frac{r^2}{\Rads^2} +1 - \frac{r_0^2}{r^2} \ .
\label{e:SAdS}
\end{equation}	
This describes a 2-parameter family of solutions, characterized by the AdS size $\Rads$ and the black hole size $\rh$, related to its mass $M$ by
\begin{equation}
r_0^2 \equiv \frac{8 \, \GN \, M}{3\, \pi} 
= \rh^2 \left( \frac{\rh^2}{\Rads^2} +1 \right)
\label{e:Mrh}
\end{equation}	
 where $\GN$ is the 5-dimensional Newton's constant.\footnote{
 Note that for notational convenience we have relabeled our coordinates: in the $\rh \to 0 $ limit we obtain global AdS \req{e:gAdS}, with the coordinate relabeling $\rho \to r$ and $\tau \to t$.
 }
 The relation \req{e:Mrh} implies that for fixed mass $\GN \, M$, the  black hole size $\rh$ is smaller than it would be in asymptotically flat case, which is consistent with what one would expect from a confining potential.
 
Once we have the bulk metric, it is a simple matter  to find the induced stress tensor on the boundary.\footnote{
One can generalize a Brown-York procedure \cite{Brown:1992br} for 
calculating a quasilocal stress tensor on a cutoff surface in terms of its mean curvature; this was done in a covariant form  in \cite{Balasubramanian:1999re} (accounting for conformal anomalies and counter-terms), 
which gave an explicit prescription.  Alternately, if one writes the metric in `\Poinc\ coordinates' generalizing \req{e:AdSPoinc}, then the stress tensor appears as the coefficient of the first subleading  (in $z$) term not specified by the boundary metric \cite{Henningson:1998gx,deHaro:2000xn}.}
  In the present situation, this is largely fixed by the symmetries:  the stress tensor must be static and homogeneous on the $S^3$ of the boundary Einstein static universe, namely
\begin{equation}
T^{\mu\nu} = \rho \, u^\mu \, u^\nu + P \, h^{\mu\nu} \ ,
\label{e:sadsTab}
\end{equation}	
where $u^\mu$ is a unit vector in the $t$ direction and $h_{\mu\nu}$ is the metric of the boundary $S^3$ (with radius $\Rads$).
This describes a perfect fluid stress tensor with energy density $\rho$ and pressure $P$.  Computing $T^{\mu\nu}$ explicitly, we find 
$\rho = 3 \, P = \frac{M}{2 \, \pi^2 \, \Rads^3}$.  Note that the stress tensor is traceless, as befits a state of the CFT, i.e.\ a conformal fluid.  
 
In the large  black hole limit, the horizon becomes planar (i.e.\ has translational symmetry), and the solution simplifies to a form akin to  the \Poinc-AdS geometry \req{e:A5S5},
\begin{equation}
ds^2 =  \frac{r^2}{\Rads^2}\left[ - \left( 1- \frac{\rh^4}{r^4} \right) dt^2
+ dx_i \, dx^i \right] + \frac{\Rads^2}{r^2}  \left( 1- \frac{\rh^4}{r^4} \right)^{\! \! -1} \, dr^2 + \Rads^2 \, d\Omega_5^2
\label{e:pSAdS}
\end{equation}	
which indeed reduces to \req{e:A5S5} for $\rh= 0$; as mentioned above, this geometry can be obtained as a near-horizon limit of near-extremal black 3-brane \cite{Maldacena:1997re}.
Although \req{e:pSAdS} is written in terms of both $\rh$ and $\Rads$, we have an additional symmetry under rescaling $r$ by $\alpha$ and simultaneously $t,x_i$ by $\alpha^{-1}$, which does not change $\Rads$ or the overall scale but only rescales the horizon radius $\rh \to \alpha \, \rh$.  Hence  \req{e:pSAdS} really describes a 1-parameter family of solutions, more naturally characterized by $\rh / \Rads$.

The \schw-AdS geometry \req{e:SAdS} can be generalized in various ways.
Thanks to the AdS asymptotics, we can have not only spherical and planar black holes, but also hyperbolic ones \cite{Emparan:1999gf}.\footnote{
In fact, one can construct horizons of arbitrary topology by quotienting by discrete isometries; this gives rise to the so-called `topological black holes'.}
Moreover, these black holes can be charged and/or rotating (see e.g.\ \cite{Emparan:2008eg} for a review).  
In fact, by virtue of being in 5 dimensions, we can even have asymptotically AdS black holes with compact horizon but non-spherical topology such as AdS black rings.  The new features compared to the asymptotically flat counter-parts come from the AdS confining potential; thus, taking the size of these black holes to be comparable to, or greater than, the AdS size $\Rads$, we get a genuinely new behavior, as manifested by the black hole thermodynamics.

\paragraph{Thermodynamics:}  
Once we know how to describe black holes in AdS, we can also elucidate their thermodynamical properties.   Note that due to the identification of the Hamiltonians, all the thermodynamic properties such as temperature, energy, entropy, etc., are in direct correspondence between the two sides.
In the CFT, a large black hole then corresponds to a hot plasma of the \GT\ degrees of freedom at the Hawking temperature. To describe black hole thermodynamics we use the usual horizon area $\sim$ entropy and surface gravity $\sim$ temperature relations.  

Since there are two length scales in the problem, $\rh$ and $\Rads$, 
the temperature (which knows both about the horizon and the asymptotics) is no longer fixed by dimensional analysis,  allowing for richer  behavior compared to the asymptotically flat case,  more akin in fact to a black hole in a box \cite{york1986black}.
In particular, the Hawking temperature of the \schw-AdS black hole \req{e:SAdS} is
\begin{equation}
T = \frac{g'(\rh)}{4  \pi}
= \frac{2\, r_+^2 + \Rads^2}{2 \pi\, r_+ \, \Rads^2} \ ,
\label{e:SAdST}
\end{equation}	
which interpolates between the asymptotically flat case for $\frac{\rh}{\Rads} \ll 1$ 
and the planar black hole case \req{e:pSAdS} for $\frac{\rh}{\Rads} \gg 1$: while  the temperature scales as inverse radius for small black holes, it {\it grows} linearly with horizon radius, 
$T \sim \frac{3}{4 \pi }\, \rh / \Rads^2$,  for large ones.
This means that a large AdS black hole has positive specific heat, so that it is thermodynamically stable: it cools off as it evaporates and therefore can be in thermal equilibrium with its Hawking radiation.  In this sense, AdS black holes are much better suited to thermodynamic analysis than  asymptotically flat ones.

Note that \req{e:SAdST} also implies that there is a minimum-temperature spherical black hole, when  $\frac{\rh}{\Rads} = \frac{1}{\sqrt{2}}$. Below this temperature, the dominant phase is described by a thermal gas of gravitons in AdS (which, having ${\cal O}(1)$ free energy, do not backreact on the pure AdS geometry).  This phase in fact dominates in the canonical ensemble till $\rh = \Rads$, above which the thermal state becomes described by the large AdS black hole.  This demarcates a first order phase transition, called the Hawking-Page transition \cite{Hawking:1982dh}.  
The \GT\ on $S^3 \times R$ can be excited to any temperature, but it likewise exhibits a corresponding phase transition, which can be viewed as a confinement-deconfinement  transition when the inverse temperature of the system becomes comparable to its size \cite{Witten:1998zw}.  At that point, the free energy jumps from ${\cal O}(1)$ 
 to ${\cal O}(N^2)$,  as the color degrees of freedom deconfine. 
Although sub-AdS size black holes with negative specific heat are not thermodynamically stable within the canonical ensemble, they are still the most entropically favorable states in the microcanonical ensemble.\footnote{
There are two caveats:  First of all, when the black hole becomes sufficiently small, it is no longer favorable for it to be smeared on the $S^5$: it becomes dynamically unstable to a Gregory-Laflamme type instability  \cite{Gregory:1993vy} and for $\rh/\Rads \approx 0.4$ it localizes on the $S^5$ \cite{Hubeny:2002xn}.  In the \GT\ this corresponds to a much more complicated state with the scalar vevs turned on.  Secondly, even  a 10-dimensional localized  black hole is not stable in the microcanonical ensemble once it becomes parametrically small, namely for $\rh/\Rads < N^{-2/17} $ \cite{Horowitz:1999uv}.
}  
However, when we keep the energy rather than the temperature fixed, the corresponding state is not a thermal state in the field theory and consequently the dual of small black holes is less well understood.

Restricting attention to a large AdS black hole, which corresponds to a thermal state in the \GT, let us now revisit the question of microscopic accounting for the black hole entropy.  
Although computing entropy of a strongly coupled hot plasma in the \GT\ is too difficult, one can at least compute the thermal entropy in Yang-Mills perturbatively and compare with the bulk entropy obtained from the black hole area.  
For a thermal gas with $N^2$ degrees of freedom at temperature $T$ in $3+1$ dimensional flat spacetime, the entropy density per unit volume scales as $\frac{1}{V} \, \Sgt \sim N^2 \, T^3$.
One might have worried that in the bulk $9+1$ dimensions already a thermal gas has entropy $\sim T^9$ which can be made arbitrarily larger than $T^3$ at high temperature, but in fact even the most entropic configuration, namely the planar black hole \req{e:pSAdS} has entropy per unit volume (in $x^\mu$ directions) 
\begin{equation}
\frac{1}{V} \, \Sbh \sim 
\frac{\Rads^2 \, \rh^3}{\gst^2 \, \lst^8} \sim 
\frac{ N^2 \, \rh^3 }{\Rads^6} 
\sim N^2 \, T^3 \ .
\label{e:entropy}
\end{equation}	
This gives another non-trivial check of the \GG\ duality: the \GT\ does indeed have (parametrically) just enough degrees of freedom to describe a higher-dimensional gravitational theory.  

To see the exact match of entropies between the bulk black hole and the thermal gas of the \GT\ is however far more difficult, because unlike in the extremal D-brane system we discussed in \sect{s:preAC}, here masses change with coupling, so the number of states computed at weak coupling does not agree with  the number of states at strong coupling in the \GT.  
If one calculates the number of states at weak coupling including the coefficient, one obtains $\Sbh = \frac{3}{4} \, {\Sgt}\mid_{\gYM=0}$ \cite{Gubser:1996de}.\footnote{
The fact that the two answers do not match precisely is not a contradiction, since they were computed in different regimes.  (We indeed expect that the strongly-coupled \GT\ answer is smaller than the weakly-coupled one, since increasing the coupling effectively raises the potential energy of the system, so there are fewer states at a fixed energy.)  Moreover, if one computes the leading corrections, the answers are consistent with smooth interpolation in $\gYM$, though why the two limits differ by such a simple factor has not been explained.}

\paragraph{Causal structure:}  
So far, we have not touched on one important feature of the  \schw-AdS solution \req{e:SAdS}: just like its asymptotically flat counterpart, its global completion has two asymptotic regions, connected by an Einstein-Rosen bridge.
The causal structure interpolates between that of \schw\ at small $r$ and AdS at large $r$, so the Penrose diagram looks similar to that of \schw, but with scri $\scri^\pm$ being timelike rather than null.\footnote{
Strictly speaking, if one fixes a conformal compactification so as to render the boundaries as straight vertical lines, one no longer has the freedom to keep the spacelike singularity drawn as straight horizontal line \cite{Fidkowski:2003nf}; instead the singularity bends inward.
}
This has an interesting implication for the dual description.  According to the picture of CFT living on the boundary of AdS, the presence of two disconnected boundaries suggests that the black hole is described by {\it two} CFTs which do not interact with each other, but are in some entangled state \cite{Maldacena:2001kr}.  This can be put on a more formal footing using the {\it thermofield} formalism \cite{israel1976thermo}, which has been very useful in trying to probe the geometry deeper.

The black hole / thermal state correspondence is however too coarse to elucidate the most intriguing quantum gravitational questions.  For this purpose we wish to understand the mapping at a much finer level.  Ultimately we hope to understand how the gauge theory `sees' the singularity, but as a first step it is useful to understand what perturbing this state in various ways does to the dual description.

\paragraph{Non-equilibrium black holes:}  
Hitherto, we have been talking about static black holes.  The easiest departure from this equilibrium context entails small perturbations.  For the conventional asymptotically flat black holes, the  evolution of a deformed black hole is  well-described by so-called quasinormal modes \cite{Kokkotas:1999bd}, specified by a discrete set of complex frequencies characteristic of the black hole (and independent of the details of the perturbation).
Perturbations of asymptotically flat black holes decay exponentially on a timescale inversely proportional to black hole mass (with a late-time power-law tail).  One can think of the perturbations as fields which either fall through the horizon or escape off to infinity.  

In AdS, the confining potential prevents fields from escaping off to infinity, but they can still decay through the horizon.  This again gives a discrete set of complex frequencies, characteristic of the AdS black hole.
Quasinormal mode frequencies for scalar perturbations were initially computed in \cite{Horowitz:1999jd}, which  confirmed that the imaginary part (corresponding to the inverse of the decay timescale)  grows linearly with temperature (equivalently $\rh$) for large black hole, but linearly with horizon area (equivalently, scattering cross-section) for small black holes.  One might be puzzled as to why an attribute of small AdS black holes does not approach the corresponding attribute of asymptotically flat black holes; the resolution here is that the attribute in question (namely its quasinormal mode frequencies) crucially hinges on the boundary conditions, and since the modes concern late-time decay, there is no causal obstruction to sensing the AdS boundary.
Interestingly, the time dependence is simpler in AdS than in asymptotically flat case:
 the decay remains  exponential at arbitrarily late times. 
 In the dual field theory, perturbing the black hole corresponds to perturbing the state away from thermal equilibrium, so quasinormal modes  characterize (e.g.\ predict the timescale for) approach to thermal equilibrium.  While this is difficult to verify by direct computation within the strongly-coupled field theory, it is consistent with expectations.

Quasinormal modes for various AdS black holes were  studied extensively (for a review see \cite{Berti:2009kk}), but perhaps the most interesting point is that for  a planar black hole \req{e:pSAdS}, there are several `massless' (or `hydrodynamic') quasinormal modes, those with arbitrarily low frequencies (and therefore arbitrarily long timescales for decay) at long wavelengths.  This will come to play an important role in our discussion in \sect{s:flugra}, where we associate them with hydrodynamic sound and shear modes of the dual conformal fluid.  
In describing the long-wavelength excitations of a black hole in terms of a boundary fluid, the fluid speed of sound one expects from \req{e:sadsTab} and conformal invariance is $v_{s} = 1/\sqrt{3}$. This might at first sight seem unlikely to emerge from the gravity side where the only natural speed is the speed of light, but remarkably, that is just what the bulk physics conspires to predict!
Actually quasinormal modes provide a great tool for studying properties of  near-equilibrium strongly coupled quantum systems; in particular we can ascertain their response and transport coefficients, which bolstered the  connections between horizon dynamics and hydrodynamics \cite{Son:2007vk}, eventually culminating in the fluid/gravity correspondence \cite{Bhattacharyya:2008jc} which we will discuss further in \sect{s:flugra}.

Of course, a far more interesting departure from equilibrium is a strongly time-dependent black hole.  One crutch to building intuition has been to use mock time-dependence by considering a static black hole in a set-up where the timelike Killing field is not manifest.  For example, one can boost a global \schw-AdS black hole\footnote{
In fact, this solution was used in \cite{Horowitz:1999gf} to construct the gravitational shock wave by taking an infinite boost limit of a global \schw-AdS black hole with fixed total energy.  In this limit the solution is no longer static.
}
 or consider it in a restricted region such as the \Poinc\ patch.\footnote{
The corresponding CFT dual on $\RR^4$ was called `conformal soliton' in \cite{Friess:2006kw}
and considerations of its causal structure \cite{Figueras:2009iu} 
(showing that while the entropy is time-independent, the area of the event horizon of the restricted solution grows and diverges at finite time) indicate that event horizon area may not be a good indicator of entropy in strongly time-dependent situations.
}
Finally, there do exist explicit genuinely time-dependent black hole solutions which retain sufficient symmetries.  The most oft-used one is Vaidya-AdS, describing an imploding spherical null `shell' (of arbitrary density profile) which collapses to a black hole.  This solution is sourced by null dust stress tensor, and interpolates between pure AdS before/inside the shell to \schw-AdS after/outside the shell.  
When the shell is infinitesimally thin, the collapse occurs maximally suddenly, so this offers a good playground for studying rapidly evolving geometries.  In the dual \GT, this describes a particular form of global `quantum quench' \cite{Calabrese07}: namely a sudden disturbance of the system (usually implemented by deforming the Hamiltonian).

From quantum gravitational standpoint, a particularly intriguing time-dependent setup is one in which a black hole collapses and then fully evaporates via Hawking radiation, for understanding the dual description completely would allow us to resolve the black hole Information Paradox.  Being described by a unitary \GT, such a bulk process would necessarily be unitary, suggesting that `information is not lost' -- but of course one also wants to see explicitly where it went (one natural guess being that it remains encoded in subtle correlations in the Hawking quanta).
  Indeed, the guarantee of unitarity provided an early example of what the \GT\ can teach us about gravity, despite the lack of computational control at strong coupling.  However, understanding the detailed process turns out to be extremely difficult.
  Since large black holes (which are homogeneous on the $S^5$) are thermodynamically stable, we need to consider sufficiently small black holes; these are not only more complicated geometrically, but also do not have a nice description in the \GT\ -- even the encoding of geometry in the vicinity of such small black holes is ill-understood due to the large non-locality of the bulk-boundary map.\footnote{
Alternately, one could consider a large black hole, but `drain away' the Hawking radiation by modifying the AdS boundary conditions from reflecting to e.g.\ transparent, or by deforming the CFT by suitable operators, thereby allowing a large black hole to evaporate. Although complete evaporation would still have to pass through a small black hole stage subject to the above-mentioned problems, to unravel the Information Paradox, one need not consider evaporating the black hole:
one can repeat a process of letting a large black hole accrete and evaporate some small fraction of its mass, say in a cyclic manner, as discussed in \cite{Marolf:2008tx}.
 However one then has to take into account the fact that one is no longer considering a closed system. 
}
  More importantly, we would also need to understand what happens {\it inside} the black hole.  This is an interesting chapter in itself, to which we turn next.
  
\paragraph{Inside the horizon:}  
So far, we have discussed physics pertaining to black hole exterior.  Since this is visible to an asymptotic bulk observer, it is not so surprising that we can describe these features in the dual \GT.
On the other hand, the most interesting stuff happens inside the horizon:  indeed, much of the motivation for considering black holes in AdS was to gain insight into the strongly quantum-gravitational effects near the singularity through the \GT\ description.  It is an important question, then, whether (and how) the \GT\ `sees' past the horizon.  Indeed, it was often suggested in the early days\footnote{
This belief has recently returned in far more concrete form with the `firewall' proposal, further mention of which we postpone till \sect{s:lessonsGR}.
} of \AC\ that the dual description stops at the horizon -- that the \GT\ cannot encode effects taking place inside the black hole.  This supposition, however, is actually more mystifying than the opposite one, since the event horizon is defined globally and thus behaves teleologically, so that the \AC\ mapping would have to be maximally temporally nonlocal in order to `know where to stop'.  It is easy to devise a simple gedanken-experiment  \cite{Hubeny:2002dg} wherein we `probe' physics inside AdS via some non-local \GT\ operator (such as a large decorated Wilson loop,  dubbed `precursor' in \cite{Polchinski:1999yd}), and {\it afterwards} excite the state in a way which in the bulk  collapses a sufficiently large black hole whose event horizon emcompasses the event which we have already probed.

While the previous argument indicates that at least in some situations the \GT\ should encode physics inside the black hole just as well as in causally trivial situations, it leaves the precise nature of such encoding obscure.  There have of course been many attempts to reconstruct the geometry (and its breakdown) inside the horizon with a variety of CFT probes,\footnote{
An early idea \cite{Balasubramanian:1999zv} was to use correlation functions for probing physics behind the horizon.  This was explicitly implemented in \cite{Fidkowski:2003nf} (see also \cite{Festuccia:2005pi}) which identified the CFT signature of the black hole singularity using analytically continued correlators of high-dimension operators. 
More direct approach  \cite{Horowitz:2009wm} used D-branes as probes, described in terms of the dynamics of rolling scalar fields in the dual \GT.
A separate set of ideas \cite{Hamilton:2006fh} is to express a local bulk operator (including those inside the horizon) in terms of boundary operators.  While this may not be possible in general, the fact that gravitational Hamiltonian is a pure boundary term allows us to relate operators probing a collapsing black hole to earlier boundary ones using boundary evolution \cite{Marolf:2008mf}.
}
but there is a wide realm left unexplored, partly due to insufficient tools to overcome the various limitations of existing techniques.  The question of understanding precisely what happens to an observer who falls into a black hole of course has wider appeal and urgency beyond the \AC\ context; but in the \AC\ setting this question is placed on a more concrete footing.  Though better understanding of the precise holographic map is essential, the amount of recent attention this problem has received invites optimism for forthcoming progress.

\subsection{Generalizations}
\label{s:generalizations}

So far, we have been describing just one particular case of the \AC\ duality, namely \req{e:ACcorr}.  There are however many ways in which the correspondence can be extended and generalized.  The earliest and most immediate one was to consider different number of dimensions, i.e.\ to describe a dual of a gravitational theory on AdS$_{d+1}$ (times a compact manifold) in terms of a $d$-dimensional CFT, on which we elaborate below.  
Another straightforward type of generalization constitutes starting with \req{e:ACcorr} and deforming both sides in a controlled way.  If we add extra terms to the Lagrangian, the \GT\ is no longer conformal; it will undergo renormalization group flow (which gives an effective description at a given energy by integrating out the higher-energy degrees of freedom).  This is directly mimicked by the behavior of the bulk geometry in the radial direction.\footnote{
One of the initial checks involved RG flows with a conformal fixed point in the IR, whose bulk dual interpolates between the original AdS asymptotics and another (smaller size) AdS, whose radius correctly accounts for the smaller number of IR degrees of freedom
\cite{Freedman:1999gp}.
In fact, in static spacetimes one can express the RG flow equation in terms of bulk Einstein's equations \cite{deBoer:1999xf}.
}  Depending on the type of deformation, we can get quite a rich set of possibilities, including ones where the  low-energy physics is confining, massive, chiral symmetry breaking, etc..  One can also replace the $S^5$ by any other Einstein manifold or a quotient of the $S^5$, which gives rise to more complicated gauge theories.
More interestingly, one can even consider different asymptotics.  

These extensions provided many further tests of the correspondence as the salient features one expects in the \GT\ were in each case faithfully mimicked by the gravity side.  
We will elaborate on a few of these developments below, primarily to exemplify the robustness of the \GG\ duality, and to give the reader some feel for how far one can wander away from the specific example \req{e:ACcorr}.  The following is certainly not meant to be an exhaustive list; numerous further examples are reviewed in \cite{Aharony:1999ti} or more recently in e.g.\ \cite{Polchinski:2010hw}.
The main point is that the \GG\ duality really refers to a broad class of dualities.  We will see in \sect{s:appliedAC}  that these allow for a rich enough structure to cover a vast arena, and in fact bear on much of everyday physics.

\paragraph{Other dimensions:}  
Generalization to other dimensions was apparent from the very outset:  \cite{Maldacena:1997re} explored a number of other examples of the \AC\ duality besides \req{e:ACcorr}.  
Instead of starting with  string theory, one can start with `M-theory' (an 11-dimensional theory whose low-energy effective action is that of 11-dimensional supergravity).  The fundamental objects of M-theory are M2-branes and M5-branes.  We can consider the low-energy limit of a stack of $N$ of these, for which much of the development indicated in \sect{s:Malda} carries through.   The near-horizon geometries that we obtain in these cases are \adss47\ with radius $\Rads = \frac{1}{2} R_{S^7} \sim \lpl N^{1/6}$ for M2-branes and \adss74  with radius $\Rads = 2 R_{S^4} \sim \lpl N^{1/3}$  for M5-branes.   M-theory on these spacetimes is then dual to a 3-dimensional conformal field theory
 (the so-called ABJM theory \cite{Aharony:2008ug}) and a 6-dimensional $(0,2)$ conformal field theory, respectively.
One can also start with string theory compactified (on $T^4$ or $K3$) down to 6 spacetime dimensions.  Take $Q_5 \gg 1$ D5-branes wrapped on the 4 compact dimensions, giving rise to a D-string in the remaining 6 dimensions, and add $Q_1 \gg 1$ D1-branes coincident with this string.  In the low-energy limit this system is described by a 2-dimensional $(4,4)$ superconformal field theory, and the near-horizon geometry of the resulting 6-dimensional black brane is \adss33\ with radii $\Rads = R_{S^3} \sim \lst\, (g_6^2 \, Q_1 \, Q_5)^{1/4}$.  Further possibilities were also mentioned in \cite{Maldacena:1997re}, involving \adss32 $\times T^6$ and even  \adss22 $\times T^6$ (the latter conjectured to be equivalent to certain quantum mechanical system).\footnote{
One should also mention another nonperturbative realization of the holographic principle, the BFSS matrix model \cite{Banks:1996vh}, which conjectures duality between M-theory in infinite momentum frame and the large-$N$ limit of a supersymmetric matrix quantum mechanics describing D0-branes.
}
Although the dual field theory is not as well understood in all these cases as for the 4-dimensional \SYM\ theory of \req{e:ACcorr}, the bulk physics is quite analogous.

One notable feature of the AdS$_3$ case (keeping the internal modes unexcited) is that the geometry remains very simple:  in 3-dimensions there are no propagating gravitational degrees of freedom, which means that any negatively curved Einstein space is locally AdS$_3$.  Nevertheless, unlike asymptotically flat 3-dimensional  spacetimes, AdS admits (the so-called BTZ) black hole solutions \cite{Banados:1992wn}, which can be obtained as quotients of AdS$_3$ but still share  many analogous properties with their higher-dimensional cousins.  Since the BTZ geometry is algebraically much simpler (and in fact, being locally pure AdS, does not receive any higher-curvature corrections\footnote{
As a consequence, the black hole entropy counting analogous to \req{e:entropy} now matches the field theory dual exactly.
}), it has provided quite a useful playground for studying features of black holes in a controlled setting.

\paragraph{Asymptotically locally AdS:}  
A relatively mild, but rather useful, generalization of the standard \AC\ correspondence is to put the field theory on a curved background different from the conformally flat ones we have been considering. 
As is well-appreciated, field theories on curved backgrounds provide a useful step towards  elucidating quantum gravity.  Even in innocuously mild contexts, we uncover many fascinating effects, such as particle production, vacuum polarization, etc..  However, technical limitations have previously restricted such studies to weakly-coupled field theories, whilst we expect further novel effects at strong coupling.  It is particularly interesting to study thermally excited states, and more ambitiously ones out of equilibrium (wherein we cannot take recourse to Euclidean techniques).  In all these regimes, \AC\ provides an excellent laboratory.  Since the background metric in \GT\ is non-dynamical (i.e., it does not need to solve any field equation, and correspondingly the field theory stress tensor does not produce any  backreaction), we have a large amount of freedom at our disposal; for example, we can put the field theory on any black hole background.  The chosen background then provides boundary conditions for the bulk dynamical spacetime, and by solving bulk Einstein's equations we can read off the requisite boundary stress tensor (see e.g.\ \cite{Fischetti:2012rd} for a detailed review), which allows us to extract the essential field theory physics; see \cite{Marolf:2013ioa} for a good overview of recent progress in this direction.

\paragraph{Other asymptotics:}  
For some, the excitement at the profound and far-reaching nature of \AC\ was tempered by the regret that \req{e:ACcorr} pertains only to asymptotically AdS bulk spacetimes.  We don't live in AdS, they remarked, but wouldn't it be nice to  describe {\it our} world holographically?
This sentiment motivated the early efforts to obtain a holographic correspondence for  spacetimes with asymptotics other than AdS.
One obvious idea is to take a further limit within the \AC\ correspondence.  The earliest attempts involved formulating flat spacetime holography (e.g.\ defining a scattering matrix) by considering processes happening on length scales much smaller than $\Rads$ \cite{Susskind:1998vk,Polchinski:1999ry,Giddings:1999jq}.
However, here we soon lose control over the \GT\ description, as anticipated from the UV/IR duality (though this has been recently revisited in e.g.\ \cite{Heemskerk:2009pn,Penedones:2010ue}, the latter recasting bulk S-matrix usefully in terms of a Mellin amplitude in CFT).

One ingenious limit which however does allow a more controlled description is the Penrose limit \cite{penrose1976any}, of zooming in on a null geodesic in the spacetime.  This generates a plane wave spacetime, considered by \cite{Berenstein:2002jq}, whose boundary is actually one-dimensional and null.
 In the \GT, this limit corresponds to a large-charge sector of the theory, which enables one to use perturbation theory on both sides and thereby retain computational control.\footnote{
The maximally supersymmetric plane wave solution of 
supergravity  \cite{Berenstein:2002jq} in fact provides the simplest solvable example of a sigma model with Ramond-Ramond background.}  In particular, one can reproduce the complete spectrum of string oscillations in this spacetime.  
While the contingency of a $d>2$ dimensional spacetime admitting only 1-dimensional conformal boundary is unusual, the plane wave spacetimes are still perfectly physically sensible geometries in terms of their causal properties: like AdS itself, they are stably causal (but not globally hyperbolic\footnote{
The fact that AdS is not globally hyperbolic is not a problem for the \AC\ correspondence, since initial conditions on any `Cauchy slice' are supplemented by reflecting boundary conditions at the boundary, rendering the Cauchy evolution in the bulk well-defined.
}).  It is interesting to note that \AC\ in fact compels us to relax causality requirements as far as possible in the causal hierarchy: one can obtain physically sensible field theories (such as those with lightlike non-commutativity considered in \cite{Hubeny:2005qu}) whose bulk dual is causal but not distinguishing.\footnote{
A spacetime is {\it causal} if it does not contain any closed causal curves, and {\it distinguishing} if distinct points have distinct causal past and future sets.  In the case studied in  \cite{Hubeny:2005qu}, all points on an entire co-dimension 1 surface have identical causal pasts and identical causal futures, despite maintaining locally Lorentzian structure.
} 
This means that although such spacetimes come arbitrarily close to having closed causal curves (and therefore one might have worried that quantum fluctuations could lead to pathological behavior), the well-posedness of the field theory dual effectively protects the bulk from causal paradoxes.

From the cosmological standpoint, it would be most desirable to obtain holography for spacetimes with positive cosmological constant.  This has motivated the so-called dS/CFT correspondence \cite{Strominger:2001pn} (see also \cite{Witten:2001kn}), which attempts to relate dynamical quantum gravity on de Sitter to a Euclidean CFT `living on' de Sitter scri.  However, although at a superficial level, de Sitter seems straightforwardly related to AdS, the different causal character of the two situations inhibits the utility of the analogy much beyond a kinematic level.  
More severely, in string theory, obtaining de Sitter scri is problematic due to quantum instability, manifested in chaotic nucleation of bubbles of metastable vacua with lower cosmological constant  \cite{Susskind:2003kw}.  As a result, the  dS/CFT correspondence is at a far less solid footing as the \AC\ correspondence. 
One can however utilize properties of a CFT for useful results bearing on cosmology, such as calculating non-gaussianities of primordial fluctuations in single field inflationary models by analytically continued 3-point functions \cite{Maldacena:2002vr}.
There have also been many efforts of embedding an inflating geometry {\it within} AdS such as \cite{Freivogel:2005qh}, but here one faces the typical problem of not having sufficient handle on the dual description in the \GT.  

 Though obtaining a 
 precise holographic duality for spacetimes with asymptotics other than AdS remains an ongoing effort, the goal of  this endeavor has mostly shifted away from trying to describe `our' universe holographically, instead focusing on elucidating the general structure of the correspondence, to understand the guiding principle behind all \GG\  dualities.

\section{Applied \AC}
\label{s:appliedAC}

As might be guessed from its breadth, the \AC\ duality has turned out  tremendously useful in elucidating both sides of the correspondence: we can use gravity to perform previously intractable calculations within strongly-coupled quantum systems on the one hand, and we can use knowledge of quantum field theories to gain insight into (quantum) gravity on the other.  Both directions are interesting from the GR perspective, since the former reveals the surprisingly vast set of applications of general relativity, and the latter teaches us something fascinating about gravity itself.

\subsection{Applying \GR\ to other real-world systems}
\label{s:flugra}

On the application front, one may of course object that \SYM\ is a conformal field theory, quite distinct from the `real-world' systems we wish to understand.  The remarkable fact is that it nevertheless provides an invaluable toy model for studying universal quantities shared by more familiar systems, as well as for exploring new classes of strongly coupled phenomena.  Applying this philosophy to quantum chromodynamics has led to the fruitful synthesis of  lattice QCD and heavy ion phenomenology with gauge/string duality via  the AdS/QCD program; 
for a concise overview see \cite{Mateos:2007ay}, and for more extensive reviews see \cite{Gubser:2009md,CasalderreySolana:2011us}.
While the potentiality of connections between particle physics and string theory had been to some extent presaged since the 70's, it seems rather more surprising that one can likewise apply this philosophy to various condensed matter systems.  This program, known as AdS/CMT, has been successful in explaining an impressive array of physical effects raging from superfluid transitions to non-Fermi liquid behavior; for extensive reviews see e.g.\ \cite{Hartnoll:2009sz,Herzog:2009xv,McGreevy:2009xe}. 
In both of these programs, the \AC\ correspondence not only provided the best (and often the only) tool to tackle the physical problems of interest, but even more remarkably, these computations actually suggest values which agree with experimentally measured quantities\footnote{
One of the early successes concern the shear viscosity of the quark-gluon plasma;  holographic arguments  \cite{Kovtun:2004de} indicate that in a certain regime (when the holographic dual corresponds to classical general relativity) the dimensionless ratio of viscosity to entropy density has a universal lower bound, $\eta/s \ge 1/4\pi$, although the bound can be violated away from this regime (cf.\ eg.\
\cite{Kats:2007mq});
gravity side (and therefore the strongly coupled ${\cal N}=4$ SYM) would saturate the bound,
while the actual value measured in quark-gluon plasma is only slightly above this lower bound. In contrast, it would diverge at weak coupling.
} in the real world! 
Both of these programs have burgeoned into active and ongoing research areas.  Below we highlight only one aspect from each to exemplify their success, referring the reader to the above-mentioned excellent reviews for a more complete overview of the subject.

\paragraph{AdS/QCD:}  
Quantum chromodynamics, the theory of the strong interactions between quarks and gluons, is a gauge theory based on the gauge group $SU(3)$ (i.e.\ quarks come in $N_c =3$ colors).  Unlike \SYM, QCD is neither supersymmetric, nor a conformal theory since the coupling runs: at large energy scales the coupling becomes weak (i.e.\ QCD is asymptotically free), whereas at low energies it is strong.\footnote{
This makes direct calculations extremely difficult, and the pre-\AC\ state of the art for low-energy calculations was achieved by lattice computations (which however are not well suited to time-dependent context).
}  Below a certain  `confinement' temperature, the physical degrees of freedom are confined into color singlet hadrons (so thermodynamic quantities scale as $N_c^0$), whereas for higher temperatures the quarks and gluons deconfine into a `quark-gluon plasma' (with $N_c^2$ scaling).  At temperatures slightly above the deconfinement transition, which are typically the relevant ones for the quark-qluon plasma created in heavy ion colliders, the quarks and gluons are still strongly coupled. This naturally provides an excellent window of opportunity for holographic techiques.
    More broadly,  while at zero temperature, \SYM\ and QCD are very different, one would expect that since finite temperature breaks both supersymmetry and conformal invariance,  the two theories become more alike in that regime, as has indeed been observed experimentally.

Understanding confinement remains one of the important problems in QCD, so it is inviting to study it in other theories which serve as toy models for the real world.
Confinement indicates that quarks are connected by flux tubes, with energy proportional to length.  In the bulk dual, this flux tube (the large-$N$ version of the QCD string) stretching between the quarks is codified by a fundamental string, which ends on the boundary at the position of the quarks.\footnote{
More accurately, since strings are oriented, its endpoints represent a quark + anti-quark pair.  The mass of these diverges with the radial position of the string endpoints, which one can however regulate by putting in additional D-brane in a suitable configuration, so as to terminate the string at finite distance radius.  
}   Hence the thick QCD string in 4 dimensions gets described by an infinitesimally thin fundamental string in the 5-dimensional bulk.  Since the fundamental string wants to minimize its worldsheet area (which determines its energy), it `hangs' form the boundary into the bulk.  We have already seen an analogous effect manifested in \fig{f:AdS} (green curves, which in this context would represent snapshots of the string at a given time, and whose endpoints would represent the quarks). 
While the SYM on $\RR^4$ does not exhibit confinement (there is no independent length scale in the problem), we have already seen in  \sect{s:BHs} 
that SYM on Einstein static universe does exhibit a confinement-deconfinement transition \cite{Witten:1998zw}.
More realistic setup involves a geometry where a spatial circle smoothly caps off, such as the `AdS soliton' constructed by \cite{Horowitz:1998ha}, where the string cannot extend beyond the bottom of the geometry. 
Such a geometry automatically enforces a regime with the confining condition of energy of quark - anti-quark pair growing linearly with their separation:
 if the string endpoints are much further separated than the scale determining where the geometry caps off, the string extends along the bottom and recovers the scaling (i.e.\ linear growth with quark separation) characteristic of confinement. In this way, the  
\AC\ correspondence  offers a simple picture of quark confinement.  

Instead of considering a meson, one could also consider just a single quark.  The bulk dual is again a fundamental string ending on the quark.  As the quark moves and accelerates, the bulk string trails behind accordingly.  One can compute its backreaction on the bulk spacetime, and from this read off the boundary stress tensor $T^{\mu\nu}$, which in turn indicates the energy-momentum distribution in the strongly-coupled plasma through which the quark propagates.  This then allows us to study interesting features such as the drag the quark experiences in moving through the plasma, its radiation due to acceleration,  the propagation and dispersion of this radiation through the plasma, and so on.\footnote{
For another nice review of these effects, see e.g.\ \cite{Chernicoff:2008sa}.
}  It is rather intriguing that the bulk string codifies both the quark (and its surrounding gluonic cloud) as well as the radiation it produces.  In fact,  cutting off the same bulk configuration at different radial distances allows us to extract the physics of a `dressed quark', including radiation damping and effects of acceleration on the surrounding gluonic cloud, in very simple way \cite{Chernicoff:2010wg}.  While from the field theory perspective these are rather complicated and sometimes puzzling effects (such as the well-known pre-acceleration effect), the bulk dual naturally provides neatly-packaged and automatically self-consistent description.

\paragraph{AdS/CMT:}  
The utility of \AC\ for studying hot quark-qluon plasmas encouraged people to explore other strongly-coupled systems.  Of immediate interest are many of the ones studied in condensed matter physics, not least because understanding  strongly-correlated systems at finite temperature is extremely challenging with presently-known condensed matter techniques once a conventional description in terms of weakly-coupled quasiparticles becomes ill-suited.  Moreover, these are experimentally accessible systems which we can control, even engineer, and some may have useful technological applications.  On the other hand, the large amount of freedom in specifying the material also suggests that we have not yet encountered all of them:  there may be novel phases with remarkable physical properties which have been hitherto overlooked.  Using a dual description is likely to systematize the enumeration of possibilities, at least in some regime, and therefore point to  new phases with novel properties which experimentalists could then look for.  In fact, in quantum optics, experimentalists are now able to design an impressive range of cold atom systems with certain specified properties \cite{greiner2008cm}.  This might even allow us to do `experimental \AC' once we design a real system with a gravitational dual.
Finally, from the quantum gravity standpoint, condensed matter physics can offer interesting toy models of emergent gravity.

Although there is no obvious analog of large $N$ in condensed matter physics, one can obtain gravitational duals of  related  systems which are close enough for many purposes.  The basic ingredients are as follows:  The dimensionality of the system (typically 3 or 2 since some layered materials are effectively 2-dimensional) determines the dimensionality of the bulk; so most studies are conducted in either 3+1 or 4+1 dimensional AdS.  As in the case of quark-gluon plasma, to describe the system at finite temperature we consider a black hole in AdS.  However, here we also typically want to keep the system at a finite chemical potential; this corresponds to charging up the black hole, which simultaneously lets us tune the temperature independently of the energy density.  To model charged condensates, one can include a charged scalar field in the bulk system, which exhibits more diverse behavior.   To `latticize' the system, we can either add periodic sources on the boundary which breaks translational invariance, or let the symmetry be broken spontaneously by an instability, resulting in `striped phases' (see e.g.\  \cite{Horowitz:2012ky,Donos:2011bh} for pioneering efforts in these directions).  Such systems naturally implement momentum dissipation and often allow for new (metallic and insulating) ground states with interesting transitions between them.
Once we have set up the gravitational system, we can analyze its properties in the bulk, which translate to the corresponding properties of the boundary dual.   For example, to extract transport properties via linear response, we study linear perturbations of the black hole. 

One nice example of the power of this technique is the holographic superconductor developed by Hartnoll, Herzog, Horowitz \cite{Hartnoll:2008vx}.  Real superconductivity occurs when Cooper pairs condense at a critical temperature $T_c$, leading to infinite DC conductivity.  While conventional superconductors are well-described by BCS theory, for the unconventional ones, such as cuprates and organics, the pairing mechanism remained obscure.
To describe the basic setup in the bulk, \cite{Hartnoll:2008vx} considered an Einstein-Maxwell-scalar system which has \RN-AdS black hole above $T_c$, but a charged black hole with a scalar hair (corresponding to the condensate) below $T_c$.  One might have naively expected a requisite no-hair theorem would rule this out, but in fact, such a contingency can indeed arise, through an instability of the \RN-AdS solution.  As demonstrated by Gubser \cite{Gubser:2008px}, the effective mass of the scalar field can become negative near the horizon, resulting in the requisite instability.

\paragraph{Fluid/gravity correspondence:}  
Let us now step back from specific systems, and consider them in a more general framework.
It is well-known that strongly-coupled quantum systems have an effective description in terms of fluid dynamics.  At sufficiently long wavelengths both the quark-gluon plasma as well as many condensed matter systems  behave like fluids.  They are described by coarse-grained variables, namely the local temperature $T$ and fluid velocity $u^\mu$ (with the usual normalization $u_\mu \, u^\mu = -1$), along with the local densities of all conserved charges.  These quantities are slowly-varying (on the microscopic scale) functions of the boundary spacetime coordinates $x^\mu$.
The dynamics is captured by the conservation of the fluid stress tensor $\nabla_\mu T^{\mu\nu} = 0$ (and any other conserved currents), supplemented by constitutive relations which allow us to express $T^{\mu\nu}$ in terms of the fluid variables.\footnote{
It is convenient to organize $T^{\mu\nu}$ in terms of `boundary derivatives' $\frac{\partial }{\partial x^\mu}$; its form in terms of the tensor structures built out of the fluid variables is fixed by symmetries, up to a finite number of undetermined coefficients (functions of $T$) at each order.  These coefficients reflect the microscopic origin of the fluid.}

 To study this system in terms of its gravitational dual, one needs to find  bulk solutions which would have the same freedom of specification built in.  Finite temperature indicates that we should consider a black hole, and the long-wavelength regime requires the black hole to be planar.  When $T$ and $u^\mu$ are independent of $x^\mu$, the requisite solution is simply a stationary black hole with temperature $T$ and horizon velocity $u^\mu$, obtained by boosting \req{e:pSAdS}.  Written more conveniently in ingoing coordinates, we have
\begin{equation}
ds^2 = -2 \, u_\mu \, dx^\mu \, dr + r^2 \left(\eta_{\mu\nu} +\frac{\pi^4 \, T^4}{r^4} \, u_\mu \, u_\nu \right) dx^\mu \ dx^\nu \ ,
\label{e:bpbh}
\end{equation}	
 which indeed gives the induced stress tensor on the boundary,
\begin{equation}
T^{\mu\nu} = \pi^4 \, T^4 \left( \eta^{\mu\nu} + 4 \, u^\mu \, u^\nu \right)
\ .
\label{}
\end{equation}	
(Although for physical fluids we expect dissipative terms as well, these would not be turned on in this stationary context.)
However, to describe a fluid out of global equilibrium, we need to promote $T$ and $u^\mu$ to functions of $x^\mu$, subject to satisfying the fluid equations.  This might seem very hard since \GR\ is a non-linear theory.  For example one can't simply replace  the parameters $T$ and  $u^\mu$ in \req{e:bpbh} by expressions with $x^\mu$-dependence.  However, in order to be describable by a fluid, the system we want to consider must be slowly varying in $x^\mu$.  This suggests recasting Einstein's equations in a derivative expansion and solving them order by order in $x^{\mu}$ derivatives, subject to regularity.  

This strategy was implemented  in  \cite{Bhattacharyya:2008jc}, explicitly up to second order in derivative expansion.  We obtain solutions whose radial dependence is solved at the fully non-linear level, written in terms of functions $T(x^\mu)$ and $u^\nu(x^\mu)$.  These must satisfy a constraint equation (obtained from the $(r \mu)$ component of Einstein's equations), which precisely reproduces the generalized Navier-Stokes equations describing the fluid.  Hence we see that 5-dimensional bulk Einstein's equations (with negative cosmological constant) contain the 4-dimensional fluid Navier-Stokes equations!  For each fluid solution we  have  a corresponding bulk black hole solution whose temperature and  horizon velocity mimics that of a fluid.  Moreover, by construction the solution remains regular well past the event horizon \cite{Bhattacharyya:2008xc}.\footnote{
Despite its global definition, the teleological nature of the event horizon is rather mild in our long-wavelength regime, allowing for explicit determination.  Moreover, the pullback of the area form on the horizon gives a natural entropy current in the boundary fluid which is guaranteed to  satisfy the 2nd law of thermodynamics by virtue of the black hole area theorem.
}  This relation is known as the fluid/gravity correspondence; for a review written for GR audience see e.g.\ \cite{Hubeny:2011hd,Hubeny:2010wp}.

Of course, the idea that a black hole horizon might resemble a fluid is not new; in particular a similar notion appears in the black hole Membrane Paradigm \cite{Price:1986yy,Damour:1978cg}.  However, in the fluid/gravity correspondence, the `membrane' lives on the boundary of the spacetime and is a perfect mirror of the entire bulk physics, not just the horizon.  
We have already seen that large \schw-AdS black holes in the bulk correspond to thermal states on the boundary, and that linearized fluctuations, described by quasinormal modes that characterize the black hole, allow us to extract various response and transport properties of the dual thermal state near global equilibrium \cite{Policastro:2001yc}.  The fluid/gravity correspondence \cite{Bhattacharyya:2008jc} takes this relation to the fully non-linear level.  We can now read-off the transport and response coefficients directly from the bulk solution; gravity in effect determines the fluid specifically.

The fluid/gravity correspondence has many useful applications.  Evidently, by geometrizing the fluid we can gain further insight into its dynamics, which despite much theoretical, experimental, and computational effort, still retains fascinating mysteries.  For example, one can try to understand turbulence using the gravitational description, further mentioned below.
At a more technical level, another intriguing consequence (in the generalized context of Maxwell-Chern-Simons charged fluid) is the appearance of a new pseudo-vector contribution to the charge current, which has been ignored by 
Landau \& Lifshitz \cite{Landau:1965uq}, but which may have potentially observable effects \cite{Son:2009tf}.  

\paragraph{Brane worlds:}  
We have touched on applications of AdS/CFT to nuclear physics, condensed matter physics, and even fluid dynamics.   Let us conclude this section by mentioning one interesting  application of AdS gravity as such, this time to grand unified model building.  Known as the `brane world' scenario, this idea was developed in parallel with the \AC\ correspondence, also using string theory.  Although logically separate,  interconnections between the two programs were soon discovered.

Our spacetime appears 3+1 dimensional.  This innocuous-sounding statement implies not just that photons seem to propagate in this many dimensions (i.e.\ at presently-accessible scales we do not directly see any extra dimensions), but also that gravitational interactions satisfy 3+1 dimensional general relativity (approximated by the Newton's $1/r^2$ force law).\footnote{
The experimental bounds on the size of extra dimensions from gravity are of course much weaker; see e.g.\ \cite{ArkaniHamed:1998nn}.
}
On the other hand, as we mentioned in \sect{s:preAC}, string theory is naturally formulated in 10-dimensional spacetime, though these dimensions need not be flat or infinite.  The conventional way of making this formulation compatible with  observation has been to compactify the extra dimensions to be smaller than the present observational bounds on them, analogously to the original Kaluza-Klein idea.  Since there is a mass gap in the KK spectrum, the zero mode of the graviton would then look 4-dimensional up to the scale given by the gap.  A new possibility arises when we consider  gauge fields living on branes, since then only gravity feels the bulk; e.g.\ two extra millimeter-sized compact dimensions would allow for unification of gravity and gauge interactions at the weak scale  \cite{ArkaniHamed:1998rs}.
However, this scenario shifts the problem of large hierarchy between the weak scale and fundamental scale of gravity to one of `large' extra dimensions.

 Soon after the advent of \AC, another option was suggested \cite{Randall:1999vf}: allowing the spacetime to have a warped product structure of AdS, as opposed to a direct product structure over the internal space, allows for large extra directions which can nevertheless remain consistent with observations.
 More specifically, while gravity (which is a manifestation of spacetime curvature) of course feels all directions, it can be {\it effectively} localized on a subspace, as a `bound state' of the higher-dimensional graviton.  Although the KK spectrum is continuous, the higher dimensional effects can remain subleading. The explicit construction in \cite{Randall:1999vf} involves taking two pieces of AdS$_5$ and gluing them together (with the junction representing a 3-brane).  This obtains a zero mode of the graviton which is localized on the brane and the massive KK mode continuum which corrects the $1/r$ gravitational potential along the brane by ${\cal O}(\Rads^2/r^3)$ effects.\footnote{
An earlier construction   \cite{Randall:1999ee} involved two branes: one with `our' matter fields and another supporting the graviton mode.
The apparent hierarchy between the weak and Planck scale is then naturally generated by the warp factor scaling exponentially with compactification radius.
}

This intriguing idea generated a large industry of various generalizations  and computations of its consequences, both in cosmology and in phenomenology.  
From the GR standpoint, there are many interesting effects for brane world black holes, whose study was initiated in \cite{Emparan:1999wa}.  To make contact with \AC, one can consider just one side of the brane, and describe the physics on the brane in terms of a cut-off field theory,  weakly-coupled to gravity.

\subsection{Lessons for \GR}
\label{s:lessonsGR}

The \AC\ correspondence has also revealed many new results pertaining to \GR\ itself.  
Although many of these could in principle have been stumbled upon without \AC, the correspondence provided the incentive or means to look in the right direction.  Here we mention a few examples.

\paragraph{New solutions:}  
We have already seen several examples wherein \AC\ guided exploration 
 leading to unexpected new solutions.\footnote{
Some can be expressed analytically, some only numerically or approximately in a perturbative expansion; but they can be analysed and studied, often revealing interesting new physics.
}    One example are the hairy black holes uncovered in the process of trying to understand superconductivity.  
Another is the large set of black hole solutions with no symmetries, arising in the fluid/gravity context (these are described in terms of lower-dimensional fluid solution, in a derivative expansion).  
Yet another example, in asymptotically locally AdS context,
  is a `flowing funnel' solution \cite{Fischetti:2012vt,Figueras:2012rb} which constructs a stationary black string whose horizon is nevertheless not a Killing horizon.\footnote{
  Although this would naively appear to violate the rigidity theorem \cite{Hollands:2006rj}, the horizon in question is non-compact, thereby evading rigidity; in the dual language, one can have entropy production even in steady state.}
While these examples all involved black holes, there have also been interesting smooth causally trivial solutions, such as the AdS  geons constructed in \cite{Dias:2011at}, which are gravitational soliton-like solutions to Einstein's equation with a helical symmetry. 

\paragraph{Instability of AdS:}  
While AdS is known to be linearly stable, the non-linear stability of AdS has been analysed only recently, cf.\ \cite{Horowitz:2014hja} for a brief review.  
The program turned out far richer than one might have anticipated.  The first surprise came with the numerical study by Bizon \& Rostworowski \cite{Bizon:2011gg}, who considered the analog of Choptuik phenomena in AdS.  One would expect that collapsing a scalar shell (specified by its amplitude $A$, with some fixed profile) in AdS will collapse to form a black hole for sufficiently large amplitude $A>A_c$  and disperse for  $A<A_c$.  Near $A \approx A_c$ we should see the usual critical behavior, at least locally.\footnote{
  Unlike the asymptotically flat case, though, one cannot characterize the final mass scaling in the critical regime as $M\sim (A-A_c)^\gamma$, since the AdS reflecting boundary conditions will cause the  initially-formed black hole to accrete the rest of the scalar field as time evolves.
}
These expectations turned out partly correct, but for the unexpected result of \cite{Bizon:2011gg} that in fact after long enough time, the black forms no matter how small $A$ is, at least for a large class of profiles.  For a range of $A$'s slighter smaller than $A_c$, the field disperses after the first implosion but collapses on the second one, for a range of smaller $A$'s the field collapses on the third implosion, and so forth, all the way down to vanishing $A$.  This implies that AdS is non-linearly unstable, in the sense of arbitrarily small perturbation evolving to a black hole.   Of course, from the field theory standpoint, this might not be so surprising, since we would expect any initial disturbance to eventually thermalize. 

 The initial expectation following this dramatic result was that any small deformation of AdS will form a black hole.  This is however not true either.   For example the geons of \cite{Dias:2011at} mentioned above provide an explicit example of a near-AdS purely gravitational configuration which does not form a black hole.  In the Einstein-scalar system, it was soon discovered that with different starting profiles, one could obtain e.g.\ time periodic solutions \cite{Maliborski:2013jca}, though the issue of characterizing the islands of stability has not been fully settled.   So it appears that while most perturbations of AdS eventually form black holes, there are small islands of stability which remain regular.  The full story is however yet to be understood.

\paragraph{Turbulence:}  
As mentioned above,
fluids display rather rich dynamics, including turbulence.
Hence one expects that, in the appropriate regime, event horizons of large black holes in AdS likewise exhibit this kind of instability.  From a conventional GR viewpoint, trained by the familiar case where a black hole dissipates perturbations maximally rapidly, 
the idea that an  event horizon could exhibit such a rich behavior may verily seem  radical.   Nonetheless, this expectation has been confirmed through numerical studies \cite{Adams:2013vsa}, which construct turbulent black holes in asymptotically AdS$_4$ spacetime by  solving Einstein equations numerically.\footnote{
An earlier work \cite{Adams:2012pj} also used \AC\ to study 2-dimensional superfluid turbulence, discovering that the superfluid kinetic energy spectrum actually obeys the Kolmogorov $-5/3$ scaling law, as it would for normal fluids in 3 dimensions.
}  They find the expected inverse cascade, but also point to a novel property of the black hole horizon:  steady-state black holes dual to $d$ dimensional turbulent flows have horizons which are approximately fractal,  with fractal dimension $d + 4/3$.

This fascinating discovery motivated further exploration of the  robustness of this behaviour away from the fluid/gravity regime.  This prompted a closer look at dynamics of perturbations of asymptotically flat black holes, which indeed revealed a new type of horizon instability for rapidly rotating Kerr black holes \cite{Yang:2014tla}.  In particular, near-extremal black holes have long-lived quasinormal modes which can excite other modes and cause them to grow exponentially for a while.  The associated energy flow exhibits an inverse cascade, analogously to turbulence for 2-dimensional fluids.  This drives the black hole toward richer angular structure, and might even have observational consequences, e.g.\ for gravitational wave signals and perhaps for spectra of accretion disks.

\paragraph{Firewalls:}  
The path towards the \AC\ correspondence was motivated in large measure by the desire to understand black holes: since string theory can resolve singularities and render the entire evolution  well-defined, it should be able to shed light on the long-standing puzzles of \QG, such as the black hole Information Paradox: how can the laws of quantum mechanics (specifically unitarity) be upheld by an evaporating black hole with (nearly) thermal Hawking radiation?  Direct approach was typically stymied by lack of control over the setup, but with the advent of the \AC\ duality, tractability of this question suddenly appeared more promising: the entire gravitational dynamics of collapsing and evaporating black hole (in AdS) is fully captured by a manifestly unitary gauge theory.  This means that black hole collapse and evaporation is described by a unitary process, so information cannot be lost.  This argument was oft presented as an example of what \AC\ can teach us using the CFT side (as opposed to the more usual AdS side) as a starting point.  

However, when the earnest effort was made at actually seeing precisely how the information is recovered, the situation turned out to be far more intricate and mystifying:  Almheiri, Marolf, Polchinski, and Sully (AMPS)  \cite{Almheiri:2012rt} realized that the following three innocuous-sounding assumptions are mutually inconstent: (i) information is not lost (i.e.\ black hole formation and evaporation process is unitary), (ii) the radiation is emitted from the region near the horizon, where low energy effective field theory valid, and (iii) the infalling observer encounters nothing unusual at the horizon (i.e.\ Einstein's equivalence principle is upheld).  Judging that relaxing the last assumption is the least radical, \cite{Almheiri:2012rt} proposed the `firewall' -- a place very near the horizon where local effective field theory description breaks down.  Naturally, this proposal generated a large amount of controversy, accompanied by a flurry of putative resolutions.\footnote{
One alternative is the earlier `fuzzball' proposal; see \cite{Mathur:2008nj} for review.  This describes black hole microstates in terms of smooth horizonless geometries (one early accomplishment being the construction of regular supersymmetric solutions of the D1-D5 system \cite{Lunin:2002iz}).
}  These were however addressed in 
\cite{Almheiri:2013hfa}, where the original AMPS argument was sharpened  (though dissenting views remain as to whether the proposed resolutions are viable), and another argument was subsequently supplied in 
\cite{Marolf:2013dba}.  To date, it is still unclear what the final resolution is.   
Nevertheless, this observation already has deep implications about the encoding of bulk geometry in the dual CFT:  whilst we understand a great deal of the spacetime outside the horizon, encoding the region inside is far more subtle.

\paragraph{Entanglement and geometry:}  
The proposals of holographic entanglement entropy \cite{Ryu:2006bv,Hubeny:2007xt} suggest a mysterious connection between entanglement and geometry.  This has been followed by 
the bold proposal of \cite{Swingle:2009bg,VanRaamsdonk:2009ar,VanRaamsdonk:2010pw} that entanglement in some sense creates geometry, and conversely, disentangling the degrees of freedom associated with different spatial regions has the geometrical effect of pinching them off from each other.  Subsequently, the even more radical notion known as ``ER=EPR"  \cite{Maldacena:2013xja} posited that the spacetime should admit `Einstein-Rosen bridges' (albeit of possibly Planckian scale) associated with any EPR pairs, present in general entangled states.  This was partly inspired by the observation that although both Einstein-Rosen bridge (or wormhole) and EPR correlations appear non-local, they do not violate causality. 
More recently, there has been mounting attention on how the geometry of an Einstein-Rosen bridge relates to information theoretic constructs.  For example, following up on the suggested connections \cite{Susskind:2013aaa} between the distance from the horizon to computational complexity, 
 \cite{Almheiri:2014lwa} relates the radial bulk direction to a measure of how well CFT representations of bulk quantum information are protected from local erasures. 

Entanglement structure seems to have some deep importance, but its precise nature is far from clear.  Even within quantum mechanics, there are many distinct measures of entanglement \cite{Vidal:2002zz,Horodecki:2009zz}, with varying degrees of computability, depending on what feature one wants to focus on.
Entanglement entropy of a given subsystem, defined as the Von Neumann entropy of a reduced density matrix for that subsystem, is well-defined given a corresponding partitioning of the Hilbert space; but for the total system being in a mixed state this quantity counts not just quantum correlations but the classical ones as well.  For instance we recover the thermal entropy of the system if we take our subsystem to be the entire system.  
Some of the other measures characterizing entanglement, such as negativity, robustness, distillable entanglement, entanglement cost, etc., may have more direct link with the genuinely quantum correlations but are often less robustly defined or less easily computable.  Nevertheless, apart from having a good thermodynamic limit, the entanglement entropy has certain `nice' properties such as strong subadditivity, and correspondingly it has a simple holographic description: an extremal surface anchored on the boundary of a given region is just about the simplest geometrical construct associated with that boundary region.\footnote{
In fact, an even simpler construct is the `causal wedge' (which only depends on the bulk causal structure) and associated quantities such as the `causal holographic information' \cite{Hubeny:2012wa}.   
However, the corresponding objects in the CFT dual are yet to be identified.
}

Finally, let us close by mentioning another intriguing recent development suggestive of the deep relation between entanglement entropy and geometry.  
Entanglement entropy obeys a `first law' which can be thought of as a  quantum generalization of the ordinary first law of thermodynamics.
  Under certain special conditions,  \cite{Faulkner:2013ica} have been able to use the holographic entanglement entropy proposal to derive   linearized Einstein equations in the bulk.  Whether one can obtain the full nonlinear Einstein's equations from known laws governing the behavior of entanglement entropy in the CFT dual remains a fascinating open problem.

\section{Epilogue: onward to quantum gravity...}
\label{s:concl}

Our story started with black holes, and black holes have been the star player throughout, appearing at various stages in different guises.  
Black hole entropy counting motivated the advances in string theory which provided the basic framework.  Extremal black branes supplied the specific context from which \AC\ was derived.  Large black holes in AdS turned out to encode the dynamics of fluids and describe a vast array of experimentally accessible systems.  Black hole information paradox has prompted the recent exploration of connections between quantum information and geometry.  
And, of course, black holes still present a crucial playground for trying to unravel \QG.

While this has been the much sought-for Holy Grail of theoretical physics, justifiably approached from many directions, the context of \AC\ correspondence seems particularly well-suited to this endeavor.  One hint is that a local formulation of \QG\ is necessarily problematic, as recently explained in \cite{Marolf:2014yga}: any gravitational theory with universal coupling to energy is charactrerized by a Hamiltonian which lives on the boundary (more precisely is a purely boundary term on shell) -- indeed, there are no local observables in quantum gravity. 
Since all bulk dynamics freezes out for such candidate local theory allowing for an effective gravitational description,  bulk gravity must then be encoded in purely boundary dynamics in the original theory.  In fact, this emergence entails further non-locality (on top of any non-locality manifest in the original dynamics), characteristic of holographic systems.  As we have seen, the \AC\ duality implements this emergence quite naturally.

From a historical perspective, it is amusing to note how much black holes had risen in prominence over the last century,
starting from presenting almost an embarrassment to general relativity, to scarcely-believed esoteric objects, to astrophysically relevant and mathematically fascinating objects, to key constructs underlying profound dualities connecting far reaches of theoretical physics, and finally to becoming virtually ubiquitous, in describing almost every-day systems.  This has been quite an amazing climb, but we're not near the summit yet, and the best is still to come!  The glimpse at profound connections between quantum information and geometry hint at more fundamental structures to be uncovered.  The resulting scientific revolution may be just around the corner, or might still take years.  At present, though, novel vistas have been revealed, and there is much to explore and relate.
Perhaps the most intriguing aspect of  \AC\ is that it interconnects so many previously disconnected and apparently disparate ideas.
In 100 years of \GR, our revelations regarding the nature of space and time  have brought us to perhaps the most exciting era to be a physicist discovering the mysteries of the Universe.


\acknowledgments 

It is a pleasure to thank Gary Horowitz, Juan Maldacena, Don Marolf, and Mukund Rangamani for useful comments on the presentation.
I would also like to thank the  Institute for Advanced Study and the Aspen Center for Physics for hospitality during early stages of this work.
I thank the STFC Consolidated Grant ST/J000426/1 and  ST/L000407/1, FQXi Grant RFP3-1334, and the Ambrose Monell Foundation for support.



\providecommand{\href}[2]{#2}\begingroup\raggedright\endgroup

\end{document}